\documentclass[12pt,a4paper]{iopart}

 \usepackage{epsfig}

\usepackage{amssymb}

\usepackage{multirow}

\begin{document}

\title[On analytic parametrizations for  proton-proton
cross-sections and asymptotia]{A study on analytic parametrizations for proton-proton
cross-sections and asymptotia}

\author{M J Menon and P V R G Silva}

\address{Universidade Estadual de Campinas - UNICAMP,
Instituto de F\'{\i}sica Gleb Wataghin \\
13083-859 Campinas, SP, Brazil}

\eads{\mailto{menon@ifi.unicamp.br}, \mailto{precchia@ifi.unicamp.br}}

\begin{abstract}
A comparative study on some representative parametrizations for the
total and elastic cross-sections as a function of energy is presented. The
dataset comprises $pp$ and $\bar{p}p$ scattering in the c.m energy 
interval 5 GeV - 8 TeV.
The parametrization for the total cross-section at low and intermediate 
energies follows the usual reggeonic structure (non-degenerate trajectories).
For the leading high-energy pomeron contribution, we consider three distinct analytic 
parametrizations: either a power ($P$) law, or a log-squared 
($L2$) law
or a log-raised-to-$\gamma$ ($L\gamma$) law, where the exponent $\gamma$ is treated as 
a real free fit parameter.
The  parametrizations are also extended to fit the elastic (integrated) cross-section data 
in the same energy interval.
Our main conclusions are the following:
(1) the data reductions with the logarithmic laws show strong dependence on the
unknown energy scale involved, which is treated here either as a free parameter
or fixed at the energy threshold;
(2) the fit results with the $P$ law, the $L2$ law (free scale)
and the $L\gamma$ law (fixed scale and exponent $\gamma$ above 2) are all consistent
within their uncertainties and with the experimental data up to 7 TeV, but they
partially underestimate the high-precision TOTEM measurement at 8 TeV;
(3) once compared with these results, the $L2$ law with fixed scale is less
consistent with the data and, in the case of a free scale, this pomeron contribution decreases
as the energy increases below the scale factor (which lies above the energy
cutoff);
(4) in all cases investigated, the predictions for the \textit{asymptotic ratio} between the
elastic and total cross-sections, within the uncertainties, do not exceed
the value 0.430 (therefore, below the black-disc limit) and the results
favor rational limits between 1/3 and 2/5.
We are led to conclude that the rise of the hadronic cross-sections at the highest
energies still constitutes an open problem, demanding further and detailed investigation.
\end{abstract}


%\submitto{\jpg}

\vspace{1.0cm}

\centerline{\textit{J. Phys. G: Nucl. Part. Phys. \textbf{40}, 125001 (2013)}}

\maketitle

%PACS: 13.85.-t Hadron-induced high- and super-high-energy interactions,
%11.10.Jj Asymptotic problems and properties,
%13.85.Lg Total cross sections.

% \newpage

\textbf{Table of Contents}

\vspace{0.3cm}

1. Introduction

\vspace{0.12cm}

2. Formalism

\ \ \ 2.1 Analytic Parametrizations for the Cross-Sections

\ \ \ \ \ 2.1.1 Low-energy Contribution

\ \ \ \ \ 2.1.2 High-energy Leading Contributions

\ \ \ 2.2 Analytic Results for the $\rho$ Parameter

\ \ \ 2.3 Summary and Notation

\vspace{0.2cm}

3. Fit Procedures and Results 

\ \ \ 3.1 Experimental Data and Fit Procedures

\ \ \ 3.2 Results of the Individual Fits to Total Cross-Section and $\rho$ Data

\ \ \ 3.3 Extensions to Elastic Cross-Section Data

\ \ \ \ \ 3.3.1 Fit and Results

\ \ \ \ \ 3.3.2 Asymptotic Ratios

\vspace{0.2cm}

4. Discussion

\ \ \ 4.1 Preliminaries

\ \ \ 4.2 Results for the Total Cross-Section and $\rho$

\ \ \ \ \ 4.2.1 $L2$ Model \textit{versus} $P$ Model

\ \ \ \ \ 4.2.2 $L2$ Model \textit{versus} $L\gamma$ Model

\ \ \ \ \ 4.2.3 Conclusions on the Fit Results

\ \ \ 4.3 Results for the Elastic Cross-Section and Asymptotia

\ \ \ \ \ 4.3.1 Conclusions on the Fit Results

\ \ \ \ \ 4.3.2  Asymptotia

\vspace{0.2cm}

5. Summary,  Conclusions and Final Remarks

\vspace{0.2cm}

Appendix A. Derivative Dispersion Relations and the Subtraction Constant

\ \ \ \ \ \ \ \ \ \ \ \ \ \ A.1 Analytic Results

\ \ \ \ \ \ \ \ \ \ \ \ \ \ A.2 Comments on the Practical Use of Derivative Dispersion Relations

\vspace{0.1cm}

Appendix B. Global Fits to Total Cross-Section and $\rho$ Data

\newpage

\section{Introduction}
\label{s1}
The last 2 years represented a very fruitful period for the physics of the strong interactions
at high energies. The advances are directly connected with
the large amount of experimental data now available mainly from the 
Brookhaven-RHIC and the CERN-LHC (which has reached 
the 8 TeV maximal energy). However, despite the recent theoretical and experimental developments
in several fronts of  nuclear and particle physics,
the \textit{soft strong interactions} 
still constitute a great challenge for QCD \cite{ssi,pred,land}.

In fact, in the absence of a pure (model-independent) QCD formalism able to predict the 
\textit{soft scattering states} (elastic and diffractive processes), phenomenology is the 
approach expected to provide information for further theoretical developments in 
the large-distance sector. Nonetheless, it is an intrinsic characteristic of
the phenomenological approach to present efficient descriptions of the experimental data
through different physical assumptions, leading, in general, to distinct physical pictures
\cite{dremin,kaspar,fiore,matthiae}. Whereas model predictions
can be selected by some novel experimental data,
phenomenological developments (additional or improved parameters) can adjust the model
to the new experimental information. In a certain sense, this feedback process seems 
to permeate the phenomenological investigation of the elastic and diffractive scattering  
in the last decades. As a consequence, a widely accepted theoretical description of
these soft hadronic processes still constitutes an open problem in the QCD context.

In the last years, the aforementioned situation has brought great expectations 
concerning the program developed by  the TOTEM 
Collaboration at the CERN-LHC, since the aim of the experiment is just the investigation of
the elastic and diffractive processes (single and double dissociation). Despite
the intrinsic experimental and technical difficulties in reaching extremely
small scattering angles, a large amount of data on several quantities has been made available 
recently. Of interest, here we quote the high-precision measurements of
total and elastic (integrated) cross-sections at 7 and 8 TeV involving
different methods and experimental conditions \cite{totem1,totem2,totem3,totem4}.

In particle collisions, the total cross-section constitutes one of the most important
physical quantities. As the sum of the elastic and inelastic cross sections, it provides fundamental 
information on the overall interaction process. In terms of the c.m. energy, the experimental data on the hadronic
total cross-section ($\sigma_{\mathrm{tot}}$) are characterized by narrow peaks (resonances)
in the region below $\sim$ 2 GeV, followed by a slow monotonic decrease as the energy
increases in the scattering state region up to $\sim$ 20 GeV. From this region on, $\sigma_{\mathrm{tot}}$
grows smoothly and monotonically up to the highest energies with
available data (see e.g. the plots in the Review of Particle Physics (RPP) by the Particle Data Group (PDG)
\cite{pdg12}, section 46).

Although the rise of $\sigma_{\mathrm{tot}}$ at high energies is an experimental fact,
the theoretical (QCD) explanation for this increase and, most importantly,
the exact energy dependence involved has been a long-standing
problem. To some extent, the theoretical difficulty can be explained by
the optical theorem since, through unitarity, it connects
$\sigma_{\mathrm{tot}}$ with the imaginary part of the forward \textit{elastic}
scattering amplitude $F$ \cite{pred}:

\begin{eqnarray}
\sigma_{\mathrm{tot}}(s) = \frac{\mathrm{Im}\,F(s,t=0)}{s},
\label{e1}
\end{eqnarray}
where $s$ and $t$ are the Mandelstam variables.
Therefore, formally, to obtain a theoretical result for $\sigma_{\mathrm{tot}}$ demands
an input for $F(s,t)$, at least at $t=0$, and that presents a challenge for 
QCD, namely a soft scattering state and here in its simplest kinematic form the elastic
channel.

In the absence of quantum field theory results for $\sigma_{\mathrm{tot}}(s)$, exclusively in terms
of quarks and gluons, the theoretical investigation has been
historically based on some general principles and formal results obtained
in different contexts. These include  Mandelstam representation, 
analyticity-unitarity-crossing concepts and axiomatic field theory leading, in general, to results 
expressed in terms of high-energy theorems and inequalities 
\cite{eden,fisher,roy,martinblois,valin}. Among them, the Froissart-Martin
bound for the total cross-section certainly plays a central role
\cite{froissart, martin1, martin2, martin3, martin4}:
\begin{eqnarray}
\sigma_{\mathrm{tot}} (s) \leq c\ln^2 \frac{s}{s_h},
\qquad
\mathrm{as}
\qquad
s \rightarrow \infty,
\label{e2}
\end{eqnarray}
where $c \leq \pi/m_\pi^2 \approx 60\, $ mb \cite{lukmar} and $s_h$ is an 
\textit{unknown constant}.
Although associated with numerical values far beyond the energies presently
accessible in experiments, the result imposes a maximum rate
of growth for the total cross-section with $s$, namely the  \textit{log-squared bound
at the asymptotic energy region}.
After the Martin derivation in the context of the axiomatic quantum
field theory \cite{martin3}, this log-squared bound, associated with unitarity, 
has played a determinant role in model constructions, aimed to treat, interpret and 
describe soft strong interactions.

On the other hand, recently (2011), Azimov has demonstrated in a formal context that,
depending on the assumed behavior for the scattering amplitude in the
\textit{non-physical region}, $\sigma_{\mathrm{tot}}$ may rise faster than
log squared of $s$ without violation of unitarity \cite{azimov1, azimov2, azimov3}.
Moreover, he has also argued that it is not obvious whether QCD can be considered an axiomatic 
field theory, since the latter demands asymptotic elementary free states and that contrasts with 
QCD confinement \cite{azimov1}.

In the phenomenological context, an operational way to investigate the energy 
dependence of $\sigma_{\mathrm{tot}}$ has been the use of different analytic parametrizations,
dictated or inspired by the analytic $S$-Matrix formalism and the Regge-Gribov theory
\cite{pred,land,sm,edenbook,collins,grib}. As we shall discuss in some detail, 
representative analytic forms include powers of $s$ and logarithmic dependencies
on $s$ (linear and quadratic forms).

In this respect, the COMPETE Collaboration completed in 2002 a broad and detailed
comparative investigation on different parametrizations, which turned out to be one
of the most comprehensive amplitude analysis for hadron scattering
\cite{compete1,compete2}.  The approach comprised several classes of analytic parametrizations
for the amplitude, different model assumptions and used all the
forward data available at that time, on $pp$, $\bar{p}p$, mesons-$p$, $\gamma p$ and $\gamma \gamma$
scattering. A detailed quantitative procedure of ranking these models by the quality of the fit
has been employed, including seven distinct statistical indicators as well as tests on different
cutoff energies and on the universality of the leading high-energy contribution.
With this ranking scheme, the log-squared parametrization, accounting for the
rise of $\sigma_{\mathrm{tot}}$ at high energies, has been selected as
the highest ranking model \cite{compete1} (a conclusion corroborated 
by subsequent works \cite{ii,bh}). After that the selected COMPETE parametrization 
became a standard reference in successive editions of the RPP by the PDG \cite{pdg12,pdg10}. 
Moreover, a remarkable result concerns the fact that 10 years later \cite{compete1}, the COMPETE 
\textit{extrapolation}
for the $pp$ total cross-section at 7 TeV showed to be in complete agreement with the
first high-precision measurement by the TOTEM Collaboration \cite{totem1}.

On the other hand, a rather intriguing result has been obtained by the PDG in the
2012 edition. The updated fit, with the log-squared COMPETE parametrization and including in the dataset the first 7 TeV TOTEM measurement of $\sigma_{\mathrm{tot}}$
\cite{totem1}, as well as the cosmic-ray data by the ARGO-YBJ Collaboration (at  $\sim$ 100 - 400 GeV) \cite{argo}
led to a data reduction not in complete agreement with the TOTEM datum. Indeed, from figure 46.10 in \cite{pdg12}, the fit result, within the uncertainty, lies below the high-precision TOTEM measurement, despite the overall reduced chi-square of 0.96.

In the period 2011 - 2012, we developed several amplitude analyses,
addressing the possibility of a different scenario for the rise of 
$\sigma_{\mathrm{tot}}$ at the highest energies \cite{fms1,fms2,ms1}. This study was motivated by the
above mentioned theoretical arguments by Azimov and was based on an analytical parametrization
introduced by Amaldi \textit{et al} in 1977, in which the exponent $\gamma$ of the leading
logarithm contribution is not fixed at 2, but treated as a free real parameter of
the fit \cite{amaldi}. This parametrization was also used by the UA4/2 Collaboration in 1993
\cite{ua42} and in this respect, Matthiae stated \cite{matthiae}

\begin{quote}
``The principal aim is to derive from the data the value of the parameter $\gamma$ which controls the 
high-energy behaviour of the cross section and to make predictions at energies above those of
the present accelerators."
\end{quote}

The analysis by Amaldi \textit{et al.} with $pp$ and $\bar{p}p$ data up to
$\sqrt{s}_{max}$ = 62 GeV indicated 
\begin{eqnarray}
\gamma = 2.10 \pm 0.10 
\nonumber
\end{eqnarray}
and the subsequent analysis
by the UA4/2 Collaboration, with data up to $\sqrt{s}_{max}$ = 546 GeV, yielded \cite{ua42}
\begin{eqnarray}
\gamma = 2.25_{-0.31}^{+0.35}.
\nonumber 
\end{eqnarray}

Although not statistically conclusive, these numerical values obtained for $\gamma$,
within the corresponding uncertainties, seem to suggest the possibility of a rise 
of $\sigma_{\mathrm{tot}}$ faster than $\ln^2 s$.
Hence, on the basis of the theoretical arguments by Azimov and these empirical results, we understood that 
to address an investigation with the Amaldi parametrization
and updated datasets (including the TOTEM data) could be a valid strategy. Our main point, 
as explicitly stated in
\cite{fms1}, has been to attempt to look for answering two questions. Is the $\ln^2s$ dependence a unique solution
describing the asymptotic rise of the total cross section? Could the data be statistically described by another
solution, rising faster (or slower) than $\ln^2s$?

However, in this respect, there is a key issue: once $\gamma$ is treated as a real free parameter
beyond the usual difficulties
associated with the nonlinearity of the fit, we are also faced with the strong
correlation among all the free parameters involved, especially an anti-correlation
between  $\gamma$ and
a high-energy scaling parameter corresponding to the unknown constant $s_h$ in equation (\ref{e2})
(both, therefore, associated with the leading high-energy contribution). As a consequence, even obtaining solutions
that are statistically consistent, we do not have uniqueness, namely we cannot provide a unique solution
but only possible statistically consistent solutions (as discussed in some detail in \cite{ms1},
section 4.2).

Despite this limitation, under a variety of different conditions, fit procedures
and datasets, we have obtained in \cite{fms1}, \cite{fms2} and \cite{ms1} several
statistically consistent solutions, indicating a rise of $\sigma_{\mathrm{tot}}$ faster
than the log-squared behavior at the LHC energy region (including all the recent
TOTEM results at 7 and 8 TeV \cite{ms1}). Moreover, and perhaps most importantly,
extension of the parametrization for $\sigma_{\mathrm{tot}}$ to fit the elastic
(integrated) cross section data, $\sigma_{\mathrm{el}}(s)$, led to asymptotic ratios 
between $\sigma_{\mathrm{el}}(s)$ and $\sigma_{\mathrm{tot}}$ below the black-disk limit
(1/2) and
consistent with the fractional
limit 1/3 \cite{fms2,ms1}.
Although answering, at least partially, the above
mentioned two questions, the lack of unique solutions gives rise to some unavoidable
critical discussions involving different points of view, as for example, those in \cite{cbh} and in \cite{rbh}.
We shall return to the above two questions in our conclusions.

In this work, as one more step in our investigation, we present and discuss the results of a \textit{comparative
study} on some representative analytic parametrizations for $\sigma_{\mathrm{tot}}$. The focus here
lies on
(1) the consequences in the fit results imposed by the recent high-precision TOTEM measurements
at 7 TeV and 8 TeV;
(2) the extension of all the parametrizations to fit the elastic  
$\sigma_{\mathrm{el}}$ data (with the extraction of the asymptotic values for the ratio $\sigma_{\mathrm{el}}/\sigma_{\mathrm{tot}}$) and
(3) discussions of the physical and phenomenological aspects involved.

As commented below, the analysis in the energy interval from 5 GeV up to
8 TeV is limited to $pp$ and $\bar{p}p$ 
data on $\sigma_{\mathrm{tot}}$, $\sigma_{\mathrm{el}}$ and
the $\rho$ parameter defined by
\begin{eqnarray}
\rho(s) = \frac{\mathrm{Re}\,F(s,t=0)}{\mathrm{Im}\,F(s,t=0)}.
\label{e3}
\end{eqnarray}
The parametrization for $\sigma_{\mathrm{tot}}$ at low energies (below $\sim$ 20 GeV)
consists of the usual Reggeon contributions associated with non-degenerate
mesonic trajectories (power law of the energy with negative exponents).
For the dominant term at high energies, responsible for the rise of $\sigma_{\mathrm{tot}}$ (pomeron contribution), 
we consider
three independent forms: either a power law of $s$ with positive real exponent, or a log-squared of $s$
or a log-raised-to-$\gamma$  of $s$ with $\gamma$ a positive real exponent.

For the reasons to be discussed (and recalled) along the paper, our focus here is on individual fits to
$\sigma_{\mathrm{tot}}$ data and the corresponding checks on the $\rho(s)$ behavior, using singly subtracted
derivative dispersion relations (DDR). However, global fits to  $\sigma_{\mathrm{tot}}$ and $\rho$ data
are also presented and discussed in an appendix. Some results and detailed discussions present in our
previous works \cite{fms1,fms2,ms1} will be referred to and summarized along the paper.

Our main conclusions are as follows.
Including in the dataset all the TOTEM measurements at 7 and 8 TeV,
the results of the fits to $\sigma_{\mathrm{tot}}$ and $\rho$ data with the power and logarithmic laws
($\gamma$ = 2 or $\gamma$ above 2) are all almost statistically
consistent within their uncertainties and with the experimental data
up to 7 TeV. All the results, however, partially underestimate the
high-precision TOTEM measurement at 8 TeV. 
In the case of the logarithmic forms, the high-energy-scale factor has
a determinant role in the physical interpretations of the data reductions.
In particular, the log-squared law with free scale leads to a decreasing pomeron contribution
as the energy increases in the physical region between the cutoff
($\sqrt{s}_{min}$ = 5 GeV$^2$) and the scale factor
($\sqrt{s}_{h} \approx$ 7 GeV$^2$). That, however, is not the case with the other laws.
In all cases investigated, the predictions for the asymptotic ratio between the
elastic and total cross sections do not exceed the value 0.430 
 within the uncertainties,
therefore, an upper bound below the black-disk limit.

The paper is organized as follows. In section \ref{s2}, we treat the formal
aspects of the analysis, displaying the analytic parametrizations for 
$\sigma_{\mathrm{tot}}(s)$ and $\rho(s)$ and discussing some phenomenological
aspects involved. 
In section \ref{s3}, the procedures and results of the fits are presented
including the extensions to the elastic cross-section data.
A detailed and critical comparative discussion of all results obtained here
is the subject of section \ref{s4}. Our conclusions and some final remarks are the 
contents of section \ref{s5}.
In appendix A, the analytical connection between $\sigma_{\mathrm{tot}}(s)$ and $\rho(s)$ is presented,
together with some comments on the practical use of DDR and 
the role of the subtraction constant. Global (simultaneous) fits to $\sigma_{\mathrm{tot}}$ and $\rho$
data are addressed in appendix B.

\section{Formalism}
\label{s2}

In this section, we introduce the analytic parametrizations to be used
in our data reductions. Although well known by experts, we shall recall
some basic physical concepts and interpretations \cite{pred,land,sm,edenbook,collins,grib,compete1,compete2}
that will help us to discuss and
discriminate the result of the fits later in section \ref{s4}.

\subsection{Analytic parametrizations for the total cross-section}
\label{s21}

In the Regge-Gribov theory, 
for $s \rightarrow \infty$, the structure of the scattering amplitude 
in the $s$-channel is determined by
its singularities in the complex $J$-plane ($t$-channel) \cite{pred,land}. Simple poles give rise
to a power-law behavior in $s$ and higher order poles results in logarithmic dependencies \cite{land,edland}:
$s\ln s$ (double pole), $s\ln^2s$ (triple pole), etc.
Through the optical theorem (\ref{e1}), these structures constitute the basic choices in the
analytic parametrizations of the hadronic total cross-section.
In this context, the behavior of the $\sigma_{\mathrm{tot}}$ data above 5 GeV 
is usually represented by two components,
associated with low-energy ($LE$) and high-energy ($HE$) contributions:
\begin{equation}
\sigma_{\mathrm{tot}}(s) = \sigma_{LE}(s) +  \sigma_{HE}(s).
\label{e4}
\end{equation}
We shall discuss each case separately.

\subsubsection{Low-energy contribution.}
\label{s211}

This term accounts for the decreasing  of $\sigma_{\mathrm{tot}}$ in the region
5 GeV $\leq \sqrt{s} \lesssim$ 20  GeV and also for the differences
between $pp$  and $\bar{p}p$ scattering.
In the Regge-Gribov theory, this contribution is associated with Reggeon exchanges,
namely the highest interpolated mesonic trajectories provided by
spectroscopic data ($t$-channel), relating Re $J$ with
the masses $M^2$ (the Chew-Frautschi plot).
The trajectories are approximately linear defining an effective slope and intercept.
The functional form for the total cross-section associated with a simple pole
consists of a power law of $s$ with exponent (related with the intercept) around - 0.5.
In its simplest and original version (Donnachie-Landshoff model \cite{dl1,dl2,dl3}), this trajectory
is degenerate, representing both $C = +1$ and $C = -1$ mesonic trajectories, namely
($a$, $f$) and ($\rho$, $\omega$), respectively. However, several amplitude analyses, 
including also both spectroscopic and scattering data, have indicated that the best
data reductions are obtained with non-degenerate trajectories \cite{cmg,ckk,dgmp,lm,lmm}.
In this case and with our previous notation \cite{fms1}, the low-energy contribution
can be expressed as

\begin{equation}
\sigma_{LE}(s) = a_1\, \left[\frac{s}{s_l}\right]^{-b_1} + 
\tau\, a_2\, \left[\frac{s}{s_l}\right]^{-b_2},
\label{e5}
\end{equation}
where $\tau$ = -1 (+1) for $pp$ ($\bar{p}p$) scattering, $s_l$ = 1 GeV$^2$ is \textit{fixed},
while $a_1$, $b_1$, $a_2$ and $b_2$ are free fit parameters. In the phenomenological
context, the parameters $a_1$ and $a_2$ are the reggeon residues (strengths) and
$b_1$  and $b_2$ are associated with the intercepts of the trajectories
(corresponding to simple poles at $J = 1 - b_1$ and $J = 1 - b_2$).

\subsubsection{High-energy leading contributions.}
\label{s212}

The  second term accounts for the rising of the total cross-section at higher energies
and for our purposes, some comments on the different parametrizations to be considered 
here are appropriate.

Up to the beginning of the 1970s, the Reggeon contributions demonstrated good agreement 
with the smooth decrease of $\sigma_{\mathrm{tot}}$ data with the energy, demanding only an 
additional constant term 
to represent the asymptotic (Pomeranchuck) limit.
However, new experimental results by the IHEP-CERN Collaboration at Serpukhov
and subsequently at the CERN-ISR indicated the
rise of $\sigma_{\mathrm{tot}}$ above $\sim$ 20 GeV.
In the absence of a mesonic trajectory able to account for this rise,
an ad hoc trajectory has been introduced, with intercept slightly greater than 1,
namely an \textit{increasing contribution with the energy}. This $C=+1$ trajectory
(to account for an asymptotic equality between $pp$ and $\bar{p}p$ scattering)
has been associated with a simple pole in the amplitude, corresponding, therefore,
to a power law in $s$. Here, this
parametrization for the total cross-section will be expressed and denoted as
\begin{equation}
\sigma_{HE}^{sp}(s) = \delta\,\left[\frac{s}{s_h}\right]^{\epsilon},
\label{e6}
\end{equation}
where $s_h$ = 1 GeV$^2$ is \textit{fixed}, $\delta$ and $\epsilon$ are free parameters 
to be fitted and the superscript $sp$ stands for
simple pole (at $J= 1 + \epsilon$).

Another possibility to explain the rise of $\sigma_{\mathrm{tot}}$ concerned the $\ln^2{s}$ behaviour,
a result already suggested in the phenomenological context even before the experimental
evidence of the rising total cross-section \cite{heisenberg,chengwu}. Based on these and
other indications \cite{bf73,luknico,kn} (perhaps also influenced by the
log-squared bound by Froissart-Martin \cite{land}), 
the higher order poles have come to take part in amplitude analyses \cite{bc,pdg92,kvw,cudell2000}.
Here, as selected by the COMPETE
Collaboration, we shall consider only the triple pole at $J = 1$, parametrized and
denoted by
\begin{equation}
\sigma_{HE}^{tp}(s) = \alpha + \beta \ln^2 (s/s_h),
\label{e7}
\end{equation}
where $\alpha$, $\beta$ and the high-energy scale $s_h$ are free parameters
of fit 
and the superscript $tp$ stands for triple pole.

At last, as commented in our introduction, we shall also consider an instrumental parametrization to address the possibility
that the exponent in the logarithm contribution might not be exactly 2. To this end, we consider the
power behaviour in $\ln{s}$ with a real exponent. This term, possibly associated with some
kind of  \textit{effective} singularity in the amplitude, will be expressed by
\begin{equation}
\sigma_{HE}^{ef}(s) = \alpha + \beta \ln^{\gamma}(s/s_h),
\label{e8}
\end{equation}
where $\alpha$, $\beta$, $\gamma$ and $s_h$ are free fit parameters 
and the superscript $ef$ stands for effective.

At this point, it is already important to stress that parametrizations 
(\ref{e6}), (\ref{e7}) and (\ref{e8}) have different mathematical structures, different regions
of validity, leading to distinct physical interpretations.
A crucial point, not usually treated in the literature, concerns the presence of the
high-energy-scale parameter $s_h$, as discussed
in what follows (and also in section 4).

The $\sigma_{HE}^{sp}(s)$ parametrization already includes
the physical condition $\epsilon >$ 0 to account for the \textit{rise} of $\sigma_{\mathrm{tot}}$
(the original or standard soft pomeron concept). As a consequence, this term \textit{increases
as the energy increases for all values of $s$}, in particular in the region above the physical threshold for
scattering states, namely $s \geq 4m_p^2$, where $m_p$ is the proton mass. 
Moreover, as a power law, the high-energy-scale factor
$s_h$ can be absorbed by the $\delta$ parameter in data reductions,
or be fixed at 1 GeV$^2$ (for dimensional reasons) as assumed here.
Therefore, this parametrization does not depend on any energy-scale factor.

That, however, is not the case with the $\sigma_{HE}^{tp}(s)$ parametrization, since it 
increases with the energy only at $s > s_h$ (it is zero at $s = s_h$ and \textit{decreases} as
$s$ increases below $s_h$). Therefore, in this case,  the strict physical interpretation of the pomeron
exchange as responsible for the rise of the cross-section depends on the value
of $s_h$ (fixed or fitted) and therefore, also on the value of the energy cutoff $\sqrt{s}_{min}$
for the data reductions.

At last, in what concerns the $\sigma_{\mathrm{HE}}^{ef}(s)$ parametrization, it is 
not defined as a real-valued function for $s < s_h$ and as a consequence cannot 
represent a physical quantity (total cross section). Therefore, in the physical 
context, equation (8) ought to be interpreted as a contribution just starting at 
$s = s_h$ with $\sigma_{\mathrm{HE}}^{ef}(s_h) = \alpha$; from this point on, it 
increases as the energy increases (as in the standard soft pomeron concept).

We shall return to these different features of the high-energy parametrizations in our discussion
of the fit results (section \ref{s4}).

\subsection{Analytic results for the $\rho$ parameter}
\label{s22}

With the parametrizations for $\sigma_{\mathrm{tot}}(s)$, the corresponding analytic
results for $\rho(s)$ can be obtained by means of singly subtracted derivative
dispersion relations (DDR). The subject is treated in some detail in appendix A, where  discussions
on the practical use of the derivative relations and the role of the
subtraction constant involved are also presented.
Here, we express the analytic results in a similar notation as
that used for the total cross-section, including a
term with the 
subtraction constant $K$ and two other additive terms ($T_{LE}(s)$ and $T_{HE}(s)$):

\begin{eqnarray}
\rho(s) &=& \frac{1}{\sigma_{\mathrm{tot}}(s)}
\left\{ \frac{K}{s} + T_{LE}(s) + T_{HE}(s) \right\}.
\label{e9}
\end{eqnarray}
The $T_{LE}(s)$ term is associated with the low-energy contribution and 
from appendix A, reads
\begin{eqnarray}
T_{LE}(s) =
- a_1\,\tan \left( \frac{\pi\, b_1}{2}\right) \left[\frac{s}{s_l}\right]^{-b_1} +
\tau \, a_2\, \cot \left(\frac{\pi\, b_2}{2}\right) \left[\frac{s}{s_l}\right]^{-b_2},
\label{e10}
\end{eqnarray}
where, as before, $\tau$ = -1 (+1) for $pp$ ($\bar{p}p$) scattering. For the
$T_{HE}(s)$ term, we have the three forms (expressed with the corresponding
superscripts $sp$,
$tr$ and $ef$):

\begin{eqnarray}
T_{HE}^{sp}(s) = \delta\,\tan \left( \frac{\pi\, \epsilon}{2}\right) \left[\frac{s}{s_h}\right]^{\epsilon},
\label{e11}
\end{eqnarray}

\begin{eqnarray}
T_{HE}^{tp}(s) = \pi \beta \ln \left(\frac{s}{s_h}\right),
\label{e12}
\end{eqnarray}

\begin{eqnarray}
T_{HE}^{ef}(s) =
\mathcal{A}\,\ln^{\gamma - 1} \left(\frac{s}{s_h}\right) +
\mathcal{B}\,\ln^{\gamma - 3} \left(\frac{s}{s_h}\right) +
\mathcal{C}\,\ln^{\gamma - 5} \left(\frac{s}{s_h}\right),
\label{e13}
\end{eqnarray}
where
\begin{eqnarray} 
\mathcal{A} = \frac{\pi}{2} \, \beta\, \gamma,  
\quad 
\mathcal{B} = \frac{1}{3} \left[\frac{\pi}{2}\right]^3 \, \beta\, \gamma\, [\gamma - 1][ \gamma - 2], 
 \nonumber \\
\mathcal{C} = \frac{2}{15} \left[\frac{\pi}{2}\right]^5 \, \beta\, \gamma\, [\gamma - 1][ \gamma - 2]
[\gamma - 3][ \gamma - 4].
\label{e14}
\end{eqnarray} 

In the last case, the third-order expansion is sufficient to ensure the convergence 
of the fit since, 
in practice (data reductions), the $\gamma$ values within the uncertainties lie below 2.5,
as we shall show in section \ref{s3}.

\subsection{Summary and notation}
\label{s23}

In what follows, as a matter of notation and for short, we shall refer to the  three
cases of \textit{analytic parametrizations} for $\sigma_{\mathrm{tot}}(s)$ by the corresponding laws associated with the
high-energy contribution, namely
\begin{description}

\item{}
\ \ \ \ \ \ \ \ \ \ $P$ model, defined by equations (\ref{e4}), (\ref{e5}) and equation (\ref{e6});

\item{}
\ \ \ \ \ \ \ \ \ \ $L2$ model, defined by equations (\ref{e4}), (\ref{e5}) and equation (\ref{e7});

\item{}
\ \ \ \ \ \ \ \ \ \ $L\gamma$ model, defined by equations (\ref{e4}), (\ref{e5}) and equation (\ref{e8}).

\end{description}

In the first case, we have
an extended Regge parametrization (the original Donnachie-Landshoff model but with non-degenerate
trajectories), in the second case, the highest-rank parametrization 
selected by the COMPETE Collaboration and in the third case, the parametrization introduced
by Amaldi \textit{et al.}.
Note that the COMPETE parametrization is a particular case of the
Amaldi parametrization for $\gamma$ = 2 fixed.
The corresponding analytic results for $\rho(s)$ are given by equations (\ref{e9}) - (\ref{e14}).

From our discussion on the high-energy leading contributions
(section \ref{s212}) and as defined above, models $P$ and $L2$ are
analytically well defined for all values of $s$ and model 
$L\gamma$ only for $s \geq s_h$. (The region where $\ln^{\gamma}(s/s_h)$
with $\gamma$ real, not integer, is a real valued function.)
The practical effect of this analytical restriction in the data
reductions will be discussed at the end of section \ref{s32}.

\section{Fit procedures and results}
\label{s3}

\subsection{Experimental data and fit procedures}
\label{s31}

Several aspects of our methodology and fit procedures have been
already presented and  discussed in our previous analyses
\cite{fms1,fms2,ms1}. In what follows, we summarize the main points
referring also to some other aspects of specific interest here.

\begin{description}

\item{1.} 
The analysis is based only on the $pp$ and $\bar{p}p$ elastic scattering data in the
energy interval from 5 GeV up to 8 TeV.
The energy cutoff, $\sqrt{s}_{min}$ = 5 GeV, is the same used in the COMPETE and PDG 
analyses \cite{pdg12,compete1,compete2,pdg10}. The restriction to
$pp$ and $\bar{p}p$ scattering means that we are dealing with only a subset
of the reactions treated by the COMPETE and PDG analyses. However, it should be noted that
$pp$ and $\bar{p}p$ scattering
correspond to the cases of largest interval in energy with available data,
giving therefore the most complete  experimental information on
particle-particle and antiparticle-particle collisions at the highest energies.

\item{2.}
The input dataset for fits concerns \textit{only accelerator data}
on $\sigma_{\mathrm{tot}}$, $\rho$ and $\sigma_{\mathrm{el}}$.
In addition to all the recent TOTEM measurements at 7 TeV (four $\sigma_{\mathrm{tot}}$
data and four $\sigma_{\mathrm{el}}$ data) and 8 TeV (one $\sigma_{\mathrm{tot}}$
datum and one $\sigma_{\mathrm{el}}$ datum) \cite{totem1,totem2,totem3,totem4},
the experimental data below this energy region have been 
collected from the PDG database \cite{pdgdata}, \textit{without any kind of data selection
or sieve procedure}. Statistical
and systematic errors have been added in quadrature.
We note that the uncertainties in the TOTEM data are systematic (not statistical),
resulting from the combination of the different contributions in quadrature and 
considering the correlations between them (table 7 in \cite{totem2} and table IV in 
\cite{totem4}).
Estimations of the $pp$ total cross-section from cosmic-ray experiments will be
displayed in the figures as illustrative results \cite{argo,akeno,fly,auger}. The TOTEM estimation
for $\rho$ at 7 TeV \cite{totem3} is also displayed as illustration.

\item{3.}
The data reductions have been performed with the objects of the class TMinuit of ROOT Framework 
\cite{root}. We have employed the default MINUIT error analysis \cite{minuit}
with the \textit{selective criteria} explained in \cite{ms1} (section 2.2.4).
In the data reductions, all the experimental points have been treated as independent,
including the cases of more than one point at the same energy.
The error matrix provides the variances and covariances associated with each free parameter,
which are used in the analytic evaluation of the uncertainty regions 
associated with the fitted and predicted
quantities (through standard error propagation procedures \cite{bev}).
In our figures, these regions will be represented by a band, delimited by the upper
and lower uncertainty extrema.
As tests of goodness-of-fit, we shall consider the chi-square per degree of freedom
($\chi^2$/DOF) and
the corresponding integrated probability, $P(\chi^2)$ \cite{bev}. The goal is not to compare or select 
fit procedures or fit results but only to check the statistical consistence of each 
data reduction in a rather quantitative way.

\item{4.}
As commented before, our main interest \textit{is not} on global (or simultaneous)
fits to $\sigma_{\mathrm{tot}}$ and $\rho$ data using dispersion relations. The main point,
as in \cite{fms1}, concerns fits to $\sigma_{\mathrm{tot}}$ data and checks on the corresponding results
for $\rho(s)$ using \textit{derivative dispersion relations}, with the subtraction constant $K$ 
as a free fit parameter
(see our discussion in appendix A.2).
Specifically, after fitting the  $\sigma_{\mathrm{tot}}$ data through equations (\ref{e4}) - (\ref{e8}),
we fix the parameters to their central values in equations (\ref{e9}) - (\ref{e14}) for
$\rho(s)$ and with only the subtraction constant $K$ as free parameter, we fit the
$\rho$ data. We shall refer to this procedure as ``individual fits to 
$\sigma_{\mathrm{tot}}$ and $\rho$ data''. Nonetheless, global fits are also treated as a complementary study
in Appendix B and will be referred to in section \ref{s4}.

\item{5.}
The nonlinearity of the fits  demands
a choice of the initial values (feedbacks) for all free parameters \cite{bev}.
Different choices have been tested and discussed in our previous analyses.
In particular, among other choices, the results of the fits in the 2010 PDG version \cite{pdg10}
have been used to initialize our parametric set in \cite{fms1} and the results of the 2012 
PDG version \cite{pdg12} in \cite{fms2} and \cite{ms1}.
Here, given the excellent agreement between the 2002 COMPETE extrapolation
and the recent TOTEM measurements at 7 and 8 TeV, we shall use the COMPETE numerical
results \cite{compete1} as initial values in our data reductions.

\item{6.} As discussed in detail in \cite{ms1} (section 4.2),
in the cases of the leading logarithm contributions ($\gamma$ = 2 or free),
the energy scale factor $s_h$ plays a central role, not only in the data reductions but 
mainly in the physical interpretations of the results. For the reasons explained there,
we also consider here two variants in the fit procedures: either  $s_h$ as a free fit parameter
or fixed to the energy threshold for the scattering states (above the resonance region),
namely $s_h = 4m_p^2$.

\item{7.}
As in \cite{fms2,ms1}, we also address the extension of the parametrizations
for  $\sigma_{\mathrm{tot}}$ to fit $\sigma_{\mathrm{el}}$ data. In this procedure, 
from the $s$-channel unitarity, the free
exponents in the leading contributions at high energies, namely
$\epsilon$ and $\gamma$, are fixed to their central fit values to $\sigma_{\mathrm{tot}}$ data.

\end{description}

In what follows, we treat the individual fits to $\sigma_{\mathrm{tot}}$ 
and $\rho$ data (section \ref{s32}) and the extensions of the parametrizations to 
$\sigma_{\mathrm{el}}$ data (section \ref{s33}). A critical discussion on all these results
is presented in section \ref{s4}.

\subsection{Results of the individual fits to total cross-section and $\rho$ data}
\label{s32}

\begin{figure}[pb]
\centering
\epsfig{file=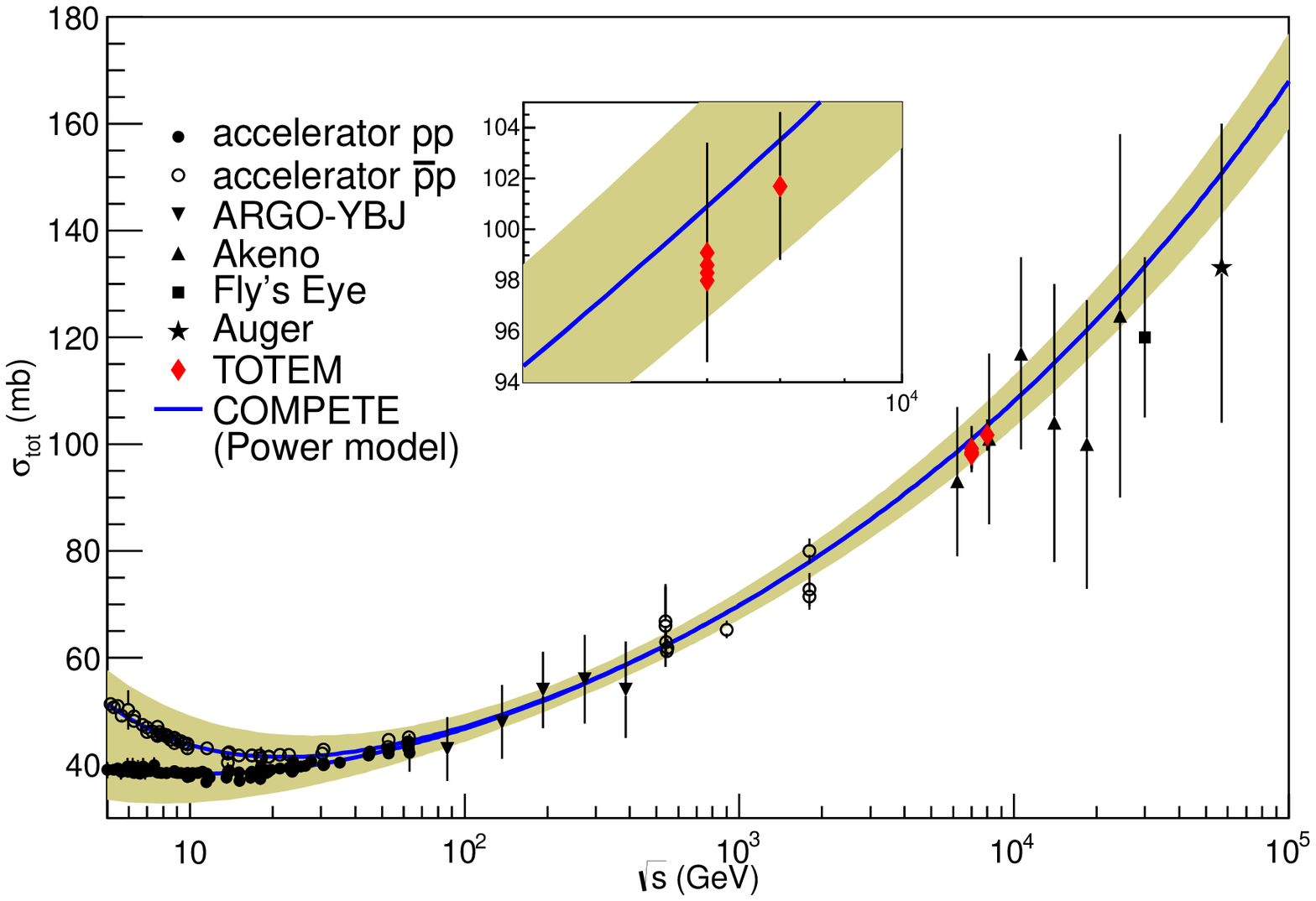,width=16cm,height=10cm}
\epsfig{file=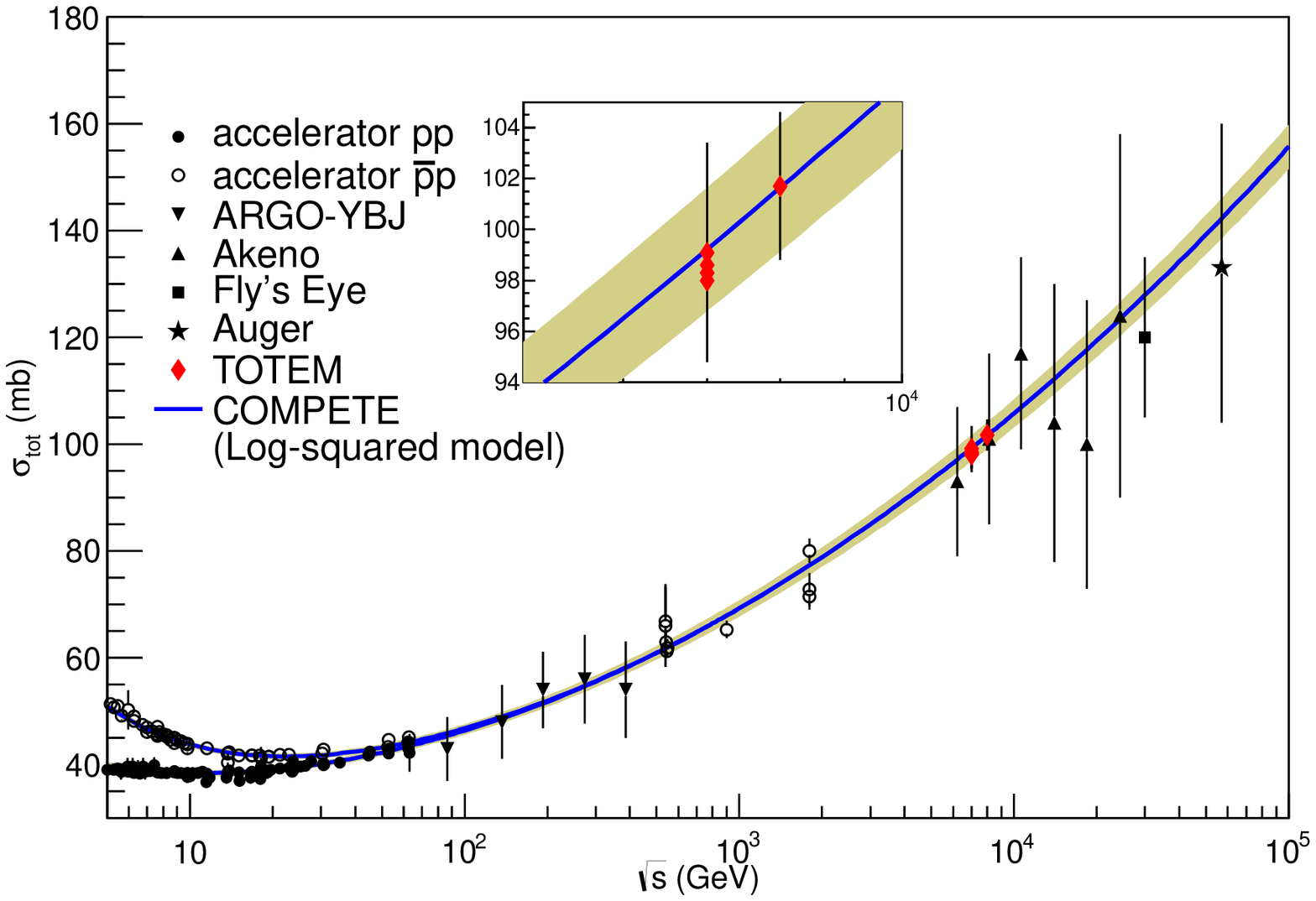,width=16cm,height=10cm}
\caption{Results for the $pp$ and $\bar{p}p$ total cross-sections from the 2002 COMPETE 
Collaboration analysis \cite{compete1}, with the parametrizations here denoted $P$ model
(equations (\ref{e4}), (\ref{e5}) and (\ref{e6})) and $L2$ model (equations (\ref{e4}), (\ref{e5}) and (\ref{e7})).
The values of the free parameters are displayed in table \ref{t1}. The references to the
accelerator data and cosmic-rays estimations are given in section \ref{s31} (item 2).}
\label{f1}
\end{figure}

To initialize our parametric set to fit the $\sigma_{\mathrm{tot}}$ data, we use the 
2002 COMPETE results, extracted from table VIII in \cite{compete1} and associated
with the models there denoted by RRE$_{nf}$
(third column in that table) and RRP$_{nf}$L2$_{u}$ (second column).
The former set applies to our $P$ model and the latter to the
$L2$ and $L\gamma$ models.
The value of the parameters, in the case of $pp$ and $\bar{p}p$ scattering
of interest here, are displayed in the second and fourth columns of our table \ref{t1}.
(The statistical information in the last lines of the table is explained in what
follows.)
The COMPETE results with these parameters and the corresponding
parametrizations, are shown in figure \ref{f1}, together with uncertainty
regions evaluated through standard propagation from the errors in the parameters
(table \ref{t1}). In the figure, it is also displayed our accelerator dataset and 
the estimations from the cosmic-rays
experiments (references in section \ref{s31}, item 2).
We shall discuss these COMPETE results together with our own fit results in section \ref{s4}. 

We note that in the case of model $L2$,
the numerical values of the COMPETE parameters reported in \cite{compete1}
(our table \ref{t1}) are not exactly the same as those reported in \cite{compete2},
which is the usual reference in the TOTEM Collaboration papers. The reason for our choice 
is the fact that in \cite{compete1}, the table provides the central values and uncertainties
for both the $P$ and the $L2$ models, which is not the case in \cite{compete2}
(where only the central values for the $L2$ model are given). Moreover, since the main role of
these parameters here is as initial values in data reductions, the small differences
in the central values are not important.

\Table{\label{t1}Fit results by the COMPETE Collaboration (2002 analysis) in the cases
of our $P$ and $L2$ models for $pp$ and $\bar{p}p$ scattering,
with the notation of equations (\ref{e4}) - (\ref{e8})
(extracted from table VIII in \cite{compete1}, models denoted RRE$_{nf}$ and RRP$_{nf}$L2$_{u}$,
respectively). The statistical information refers to the output of the first
MINUIT run, corresponding to the central values of the parameters and the dataset used here (see text).
The parameters $a_1$, $a_2$, $\alpha$, $\beta$ and $\delta$ are in mb, $s_h$ in GeV$^2$, 
$b_1$,  $b_2$,  $\gamma$ and  $\epsilon$ are dimensionless ($s_l$ = 1 GeV$^2$).} 
\begin{tabular}{c c | c c}\hline
\multicolumn{2}{c|}{$P$ model} & \multicolumn{2}{|c}{$L2$ model} \\
\multicolumn{2}{c|}{(RRE$_{nf}$ in \cite{compete1})} & \multicolumn{2}{|c}{(RRP$_{nf}$L2$_{u}$ in \cite{compete1})} \\
\hline
$a_1$        & 66.1   $\pm$ 1.2        & $a_1$             & 42.1   $\pm$ 1.3     \\
$b_1$        & 0.3646 $\pm$ 0.095      & $b_1$             & 0.467  $\pm$ 0.015   \\
$a_2$        & 35.3   $\pm$ 1.6        & $a_2$             & 32.19  $\pm$ 0.94    \\
$b_2$        & 0.5580 $\pm$ 0.0099     & $b_2$             & 0.5398 $\pm$ 0.0064  \\
$\delta$     & 18.45  $\pm$ 0.41       & $\alpha$          & 35.83  $\pm$ 0.40    \\
$\epsilon$   & 0.0959 $\pm$ 0.0021     & $\beta$           & 0.3152 $\pm$ 0.0095  \\
$s_h$        & 1 (fixed)               & $\gamma$          & 2 (fixed)            \\
             &                         & $s_h$             & 34.0   $\pm$ 5.4     \\
\hline
DOF          & 162                     & DOF               &    161               \\
$\chi^2$/DOF & 7.13                    & $\chi^2$/DOF      &    1.01              \\
$P(\chi^2)$  & 2.57$\times$10$^{-149}$ & $P(\chi^2)$       &    0.442             \\
\hline
Figure:      &       \ref{f1}          &     Figure:       &        \ref{f1}             \\
\hline
\end{tabular}
\endTable

Differently from the COMPETE analysis, our ensemble consists of the $\sigma_{\mathrm{tot}}$
data from $pp$ and $\bar{p}p$ scattering in the energy interval 5 GeV - 8 TeV.
For models $P$, $L2$ and $L\gamma$, we use as initial values the corresponding central values displayed in
table \ref{t1}. The first MINUIT run yields the $\chi^2$ for that ensemble
and central values of the parameters; the final  convergent run provides our fit
result for that ensemble and model considered.
The statistical information obtained in the first MINUIT run for the models $P$ and $L2$
(namely the COMPETE results with our ensemble) are displayed in the last 
lines of table 1.

For each model, after fitting the $\sigma_{\mathrm{tot}}$ data, we check the results for
$\rho(s)$. In this case, we fix the resulting values of the parameters to their central
values in equations (\ref{e9}) - (\ref{e14}) for $\rho(s)$ and then, with only the subtraction
constant as a free parameter, we fit the $\rho$ data.

With this procedure our fit results with the $P$ model are displayed in table
\ref{t2} (second and third columns) and in figure 2, together with the evaluated uncertainty
regions and experimental information.

\Table{\label{t2}Fit results with the $P$ model: individual fits to $\sigma_{\mathrm{tot}}$ and
 $\rho$ data and extentions to $\sigma_{\mathrm{el}}$ data. Units as in table \ref{t1}.}
\begin{tabular}{c| c| c |c}\hline
               & $\sigma_{\mathrm{tot}}$ data  &   $\rho$ data         &  $\sigma_{\mathrm{el}}$ data \\
\hline
  $a_1$        &    64.4 $\pm$ 1.8    &          64.4         & 72.4 $\pm$ 7.6\\
  $b_1$        &   0.364 $\pm$ 0.012  &         0.364         & 0.811 $\pm$ 0.028\\
  $a_2$        &    33.0 $\pm$ 2.3    &          33.0         &      0 (fixed)\\
  $b_2$        &   0.539 $\pm$ 0.015  &         0.539         &    $-$      \\
  $\delta$     &   18.94 $\pm$ 0.35   &         18.94         &  3.685 $\pm$ 0.021\\
  $\epsilon$   &  0.0926 $\pm$ 0.0016 &         0.0926        &   0.0926 (fixed)\\
  $K$          &         $-$          &     24.0 $\pm$ 4.9    &   $-$  \\
\hline
  DOF          &           162        &           75          &        105\\
  $\chi^2$/DOF &           0.91       &          1.46         &        3.49\\
  $P(\chi^2)$  &          0.794       & 5.91$\times$10$^{-3}$ & 1.08$\times$10$^{-30}$\\
\hline
Figure:        &        \ref{f2}      &          \ref{f2}     & \ref{f6}      \\
\hline
  \end{tabular}
\endTable

\begin{figure}[pb]
%\centering
\epsfig{file=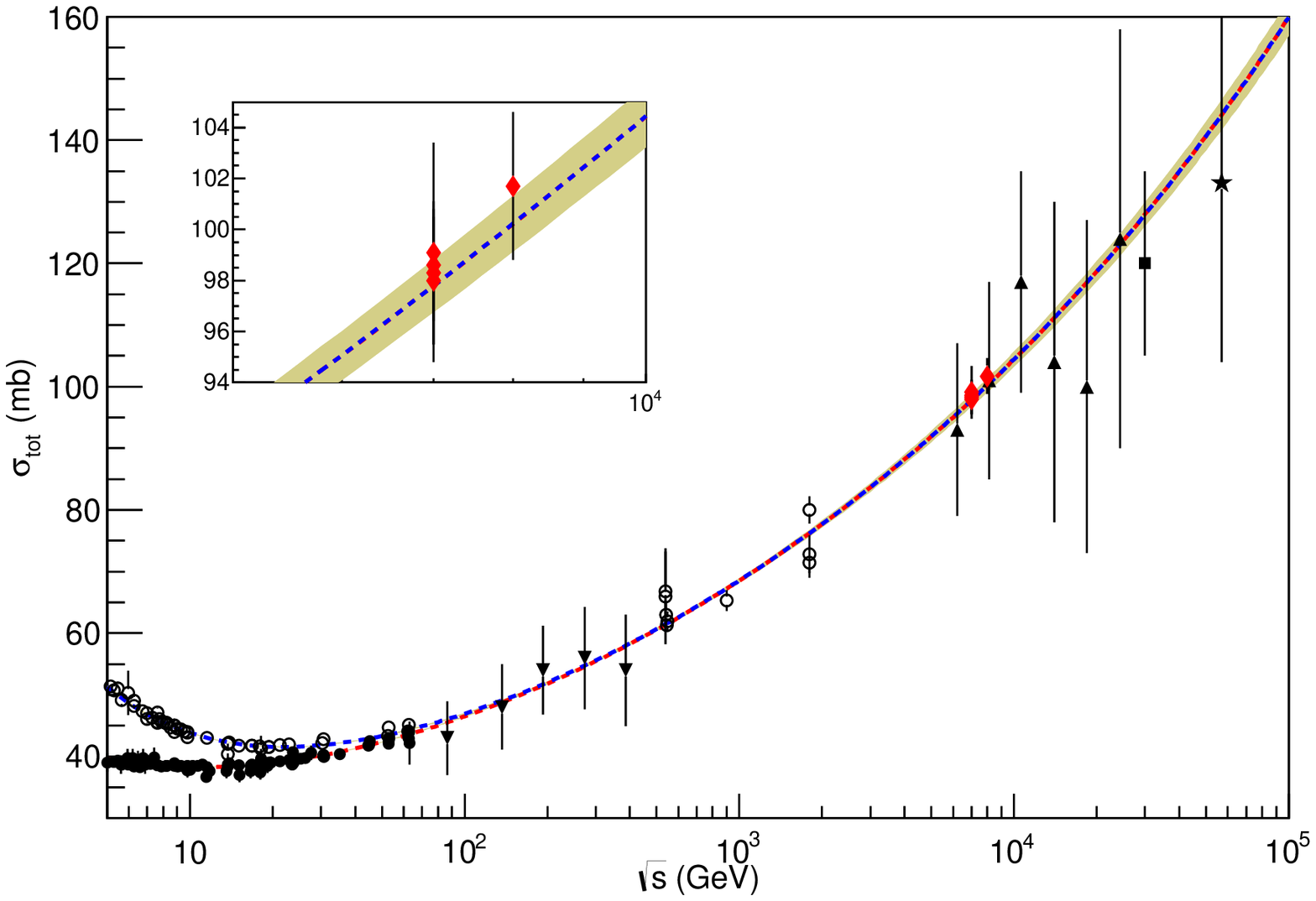,width=15cm,height=8cm}
\epsfig{file=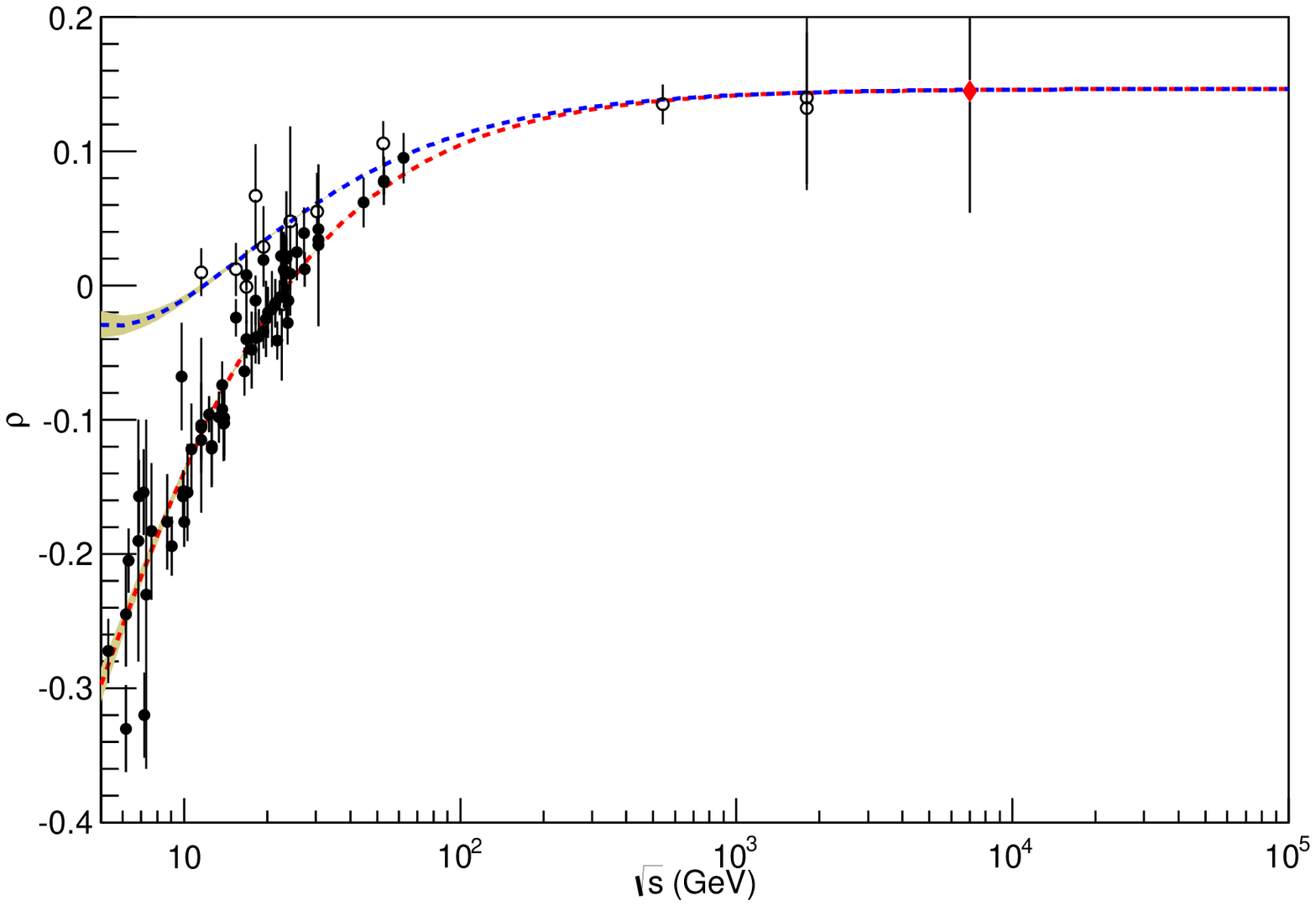,width=15cm,height=8cm}
\caption{Results of the individual fits to $\sigma_{\mathrm{tot}}$ and $\rho$ data 
with the $P$ model (table \ref{t2}). Legend on data as in figure \ref{f1}.}
\label{f2}
\end{figure}

\begin{figure}[pb]
\centering
\epsfig{file=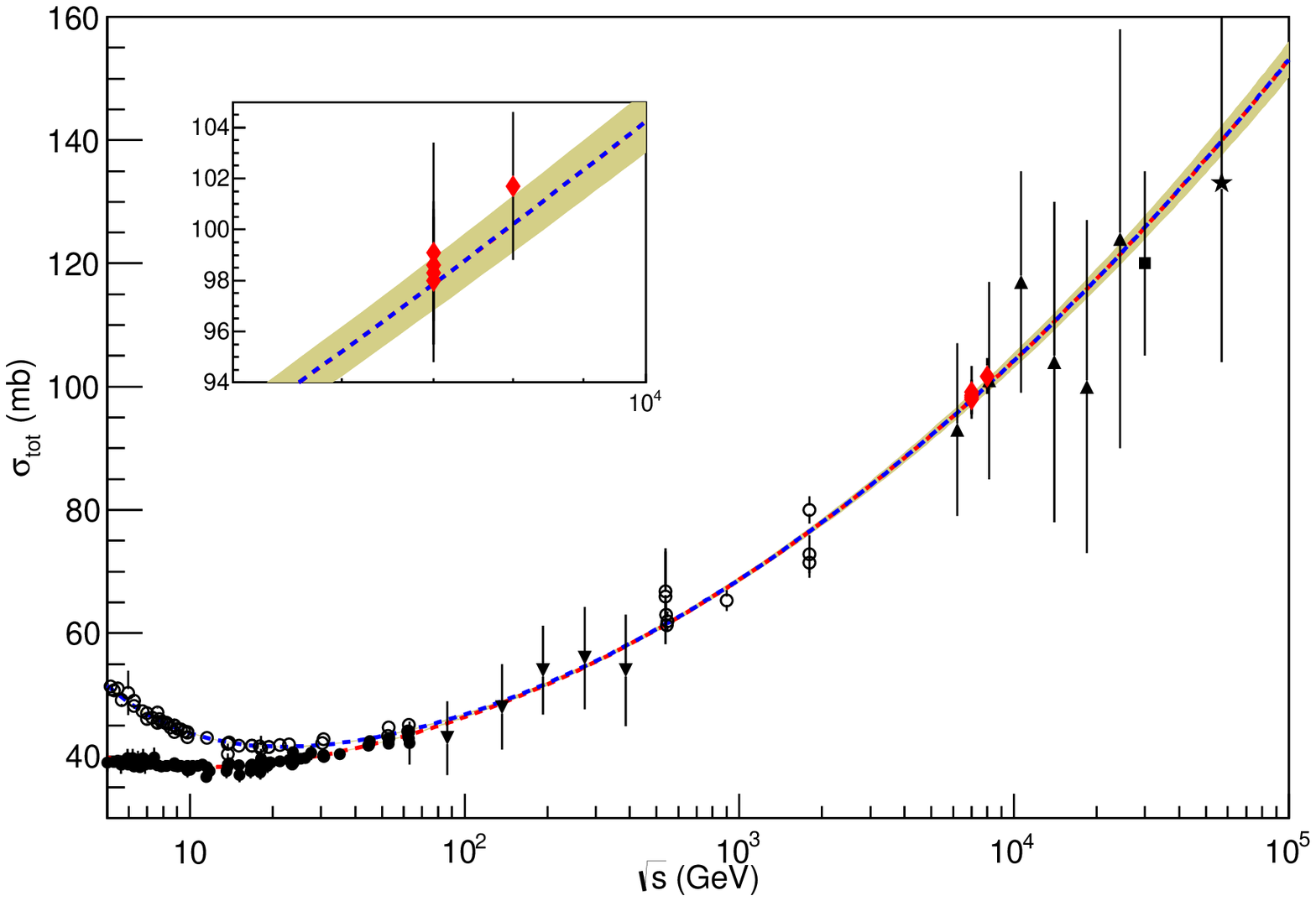,width=15cm,height=8cm}
\epsfig{file=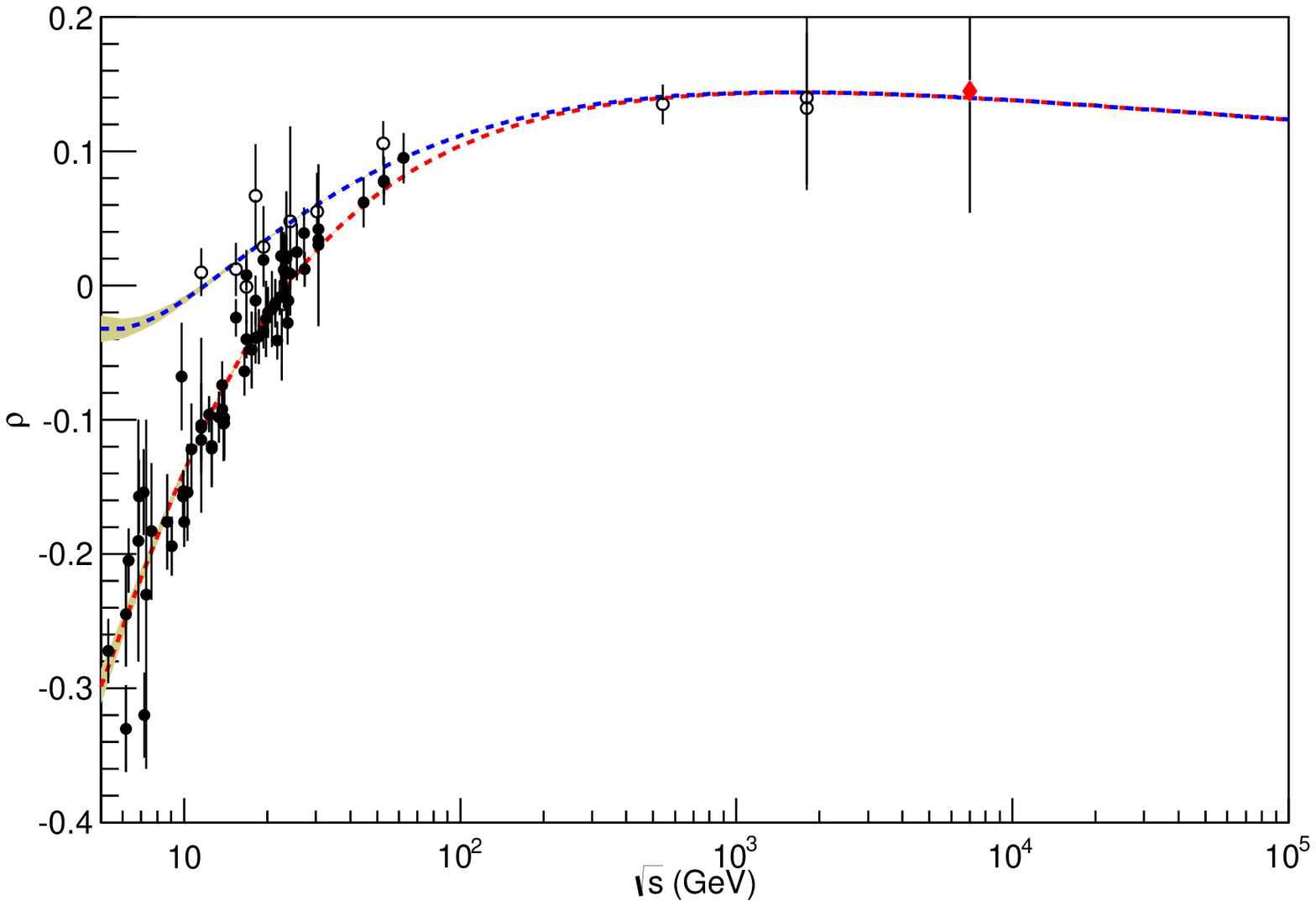,width=15cm,height=8cm}
\caption{Results of the individual fits to $\sigma_{\mathrm{tot}}$ and $\rho$ data 
with the $L2$ model and $s_h$ free (table \ref{t3}).
Legend on data as in figure \ref{f1}.}
\label{f3}
\end{figure}

\begin{figure}[pb]
\centering
\epsfig{file=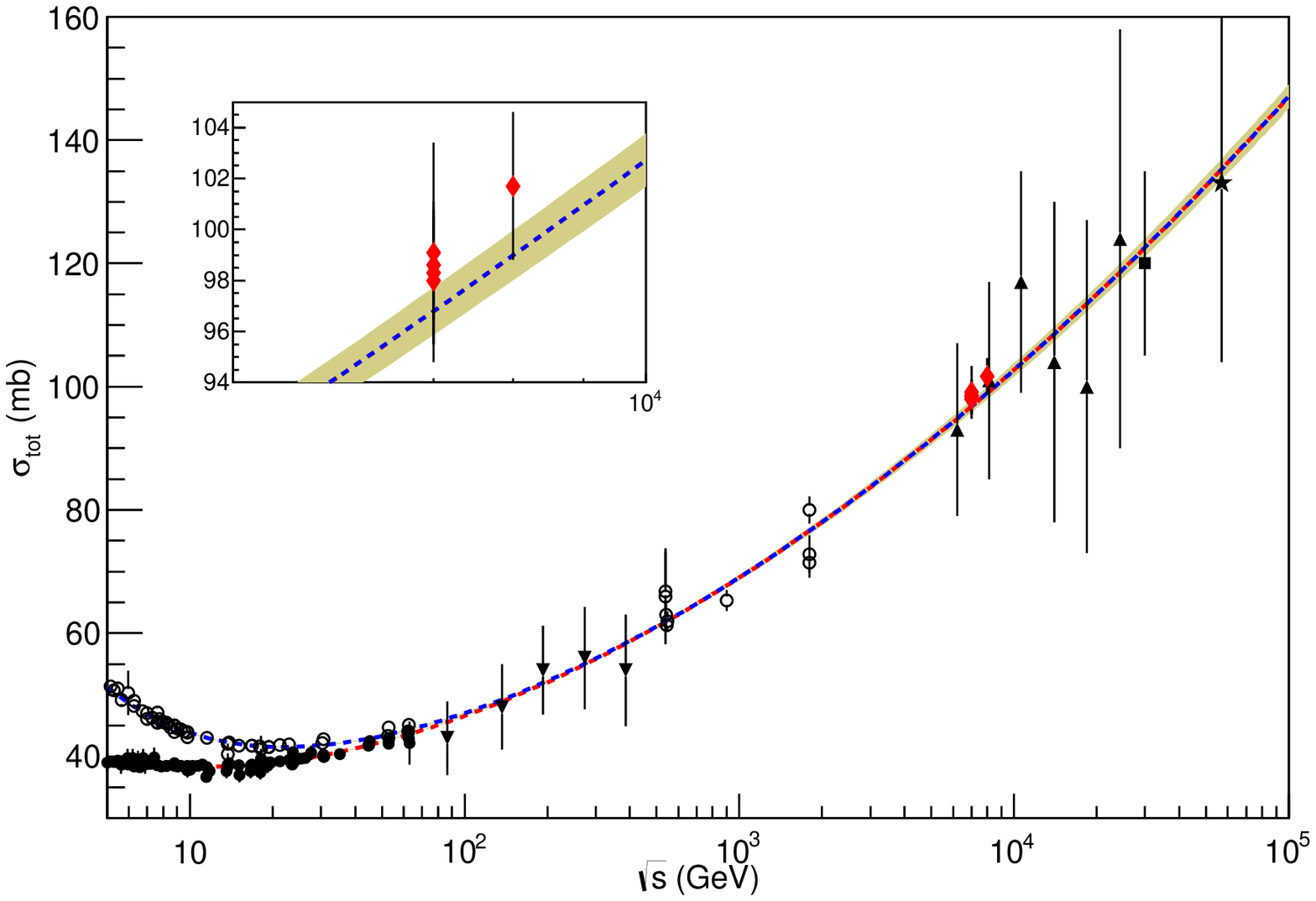,width=15cm,height=8cm}
\epsfig{file=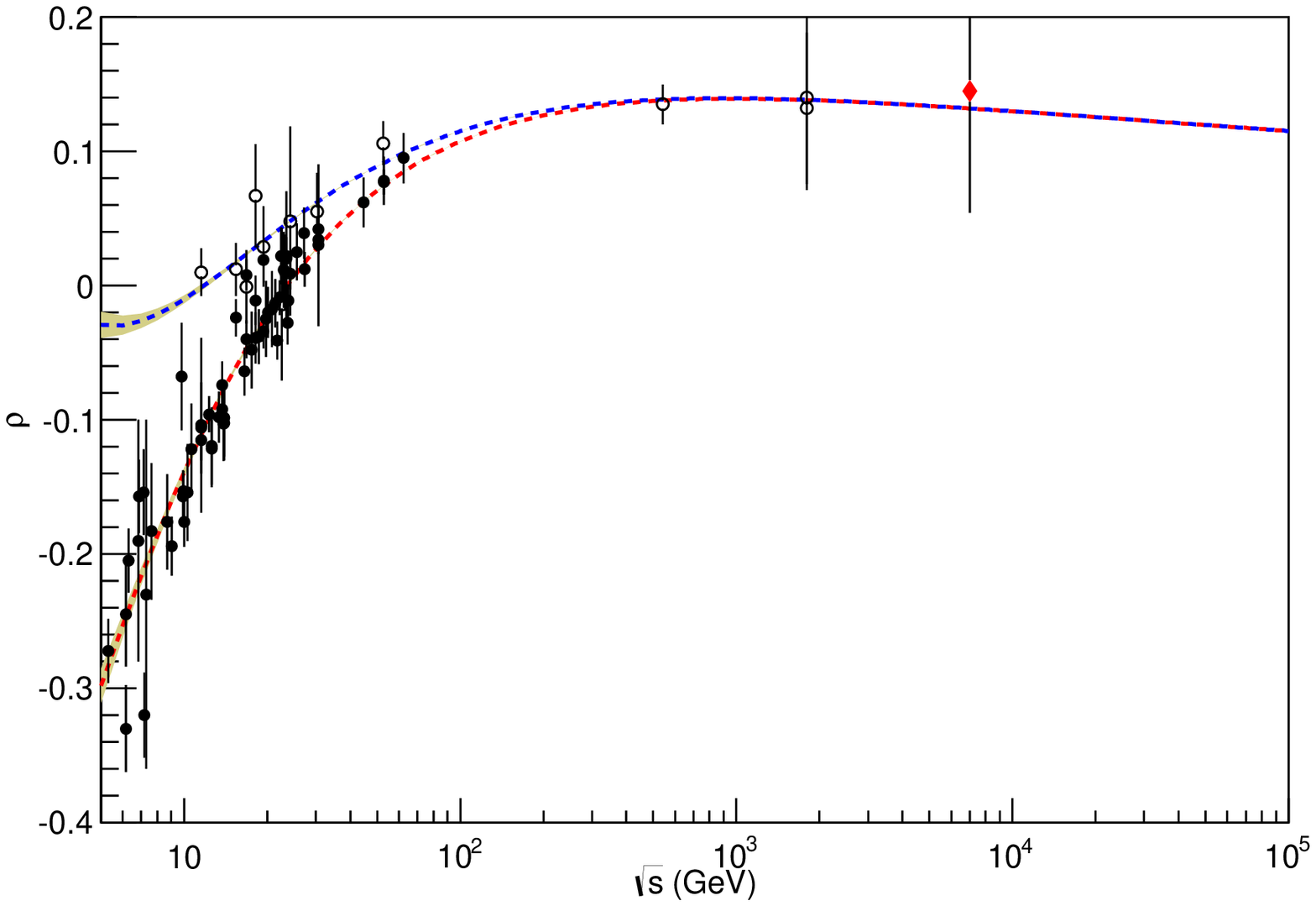,width=15cm,height=8cm}
\caption{Results of the individual fits to $\sigma_{\mathrm{tot}}$ and $\rho$ data 
with the $L2$ model and $s_h = 4m_p^2$ fixed (table \ref{t4}).
Legend on data as in figure \ref{f1}.}
\label{f4}
\end{figure}

In the case of the $L2$ model, we consider two variants,
either the high-energy-scale factor $s_h$ as a free parameter or fixed
to the energy threshold $s_h = 4m_p^2$. The results with the former variant are displayed
in table \ref{t3} (second and third columns) and figure \ref{f3} and those with the latter 
variant are shown in table \ref{t4} (second and third columns) and figure \ref{f4}.

Although the same two variants had been considered with the $L\gamma$ model,
we did not obtain full convergence in the case of $s_h$ as a free fit parameter,
but only for $s_h = 4m_p^2$ fixed. We understand that this effect can be
explained as follows.

As discussed in section \ref{s23}, model $L\gamma$ is well defined  only
for $s \geq s_h$. In practice (data reductions through the MINUIT code),
if this condition is not satisfied in the physical region above the cutoff,
the fit does not converge (because $\ln^{\gamma}(s/s_h)$ is not a real-valued 
function). 
Now, from table \ref{t3}, in the case of the
$L2$ model with $s_h$ free, we see that the fit value of this parameter
is somewhat large, $s_h \sim$ 40 GeV$^2$, lying, therefore, in the physical region of
the data reduction, above the cutoff $s_{min}$ = 25 GeV$^2$. Due to the
strong anti-correlation between the fit parameters $\gamma$ and $s_h$, 
the data reduction favors larger
values of $s_h$ which can reach the physical region (beyond the cutoff), 
where the $\ln^{\gamma}(s/s_h)$ term is not defined, resulting, therefore,
in no convergence. We understand that this trend of the data reductions
is also connected with the rather large value of the high-precision
TOTEM measurement at 8 TeV. In fact, as we have shown in \cite{ms1},
if this point is not included in the dataset (the $\sqrt{s}_{max}$ =
7 TeV Ensemble in \cite{ms1}), full convergence is obtained and $s_h$
as free parameter lies below the cutoff $s_{min}$ = 25 GeV$^2$.
However, once included in the dataset, no convergence is obtained.

\begin{figure}[pb]
\centering
\epsfig{file=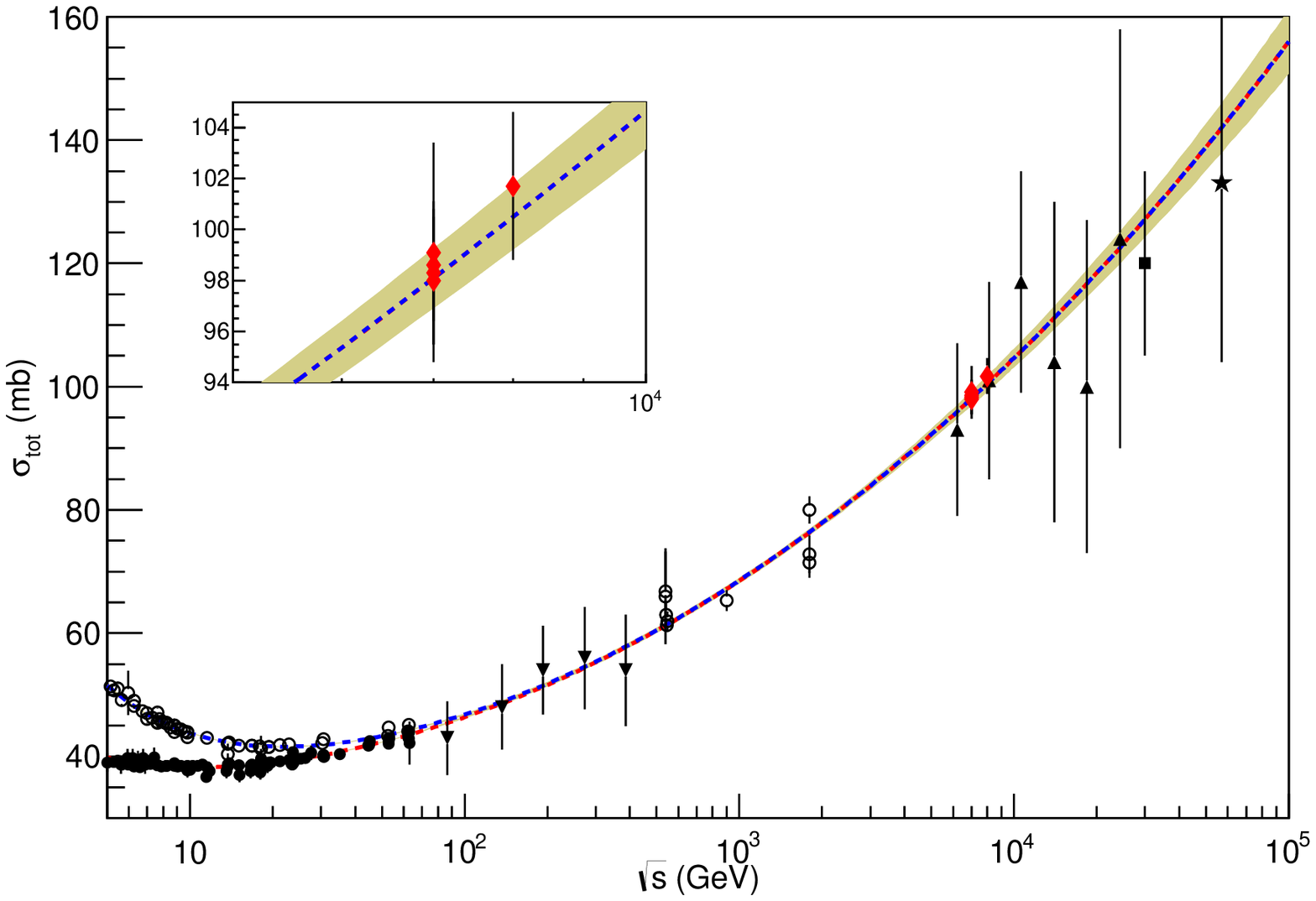,width=15cm,height=8cm}
\epsfig{file=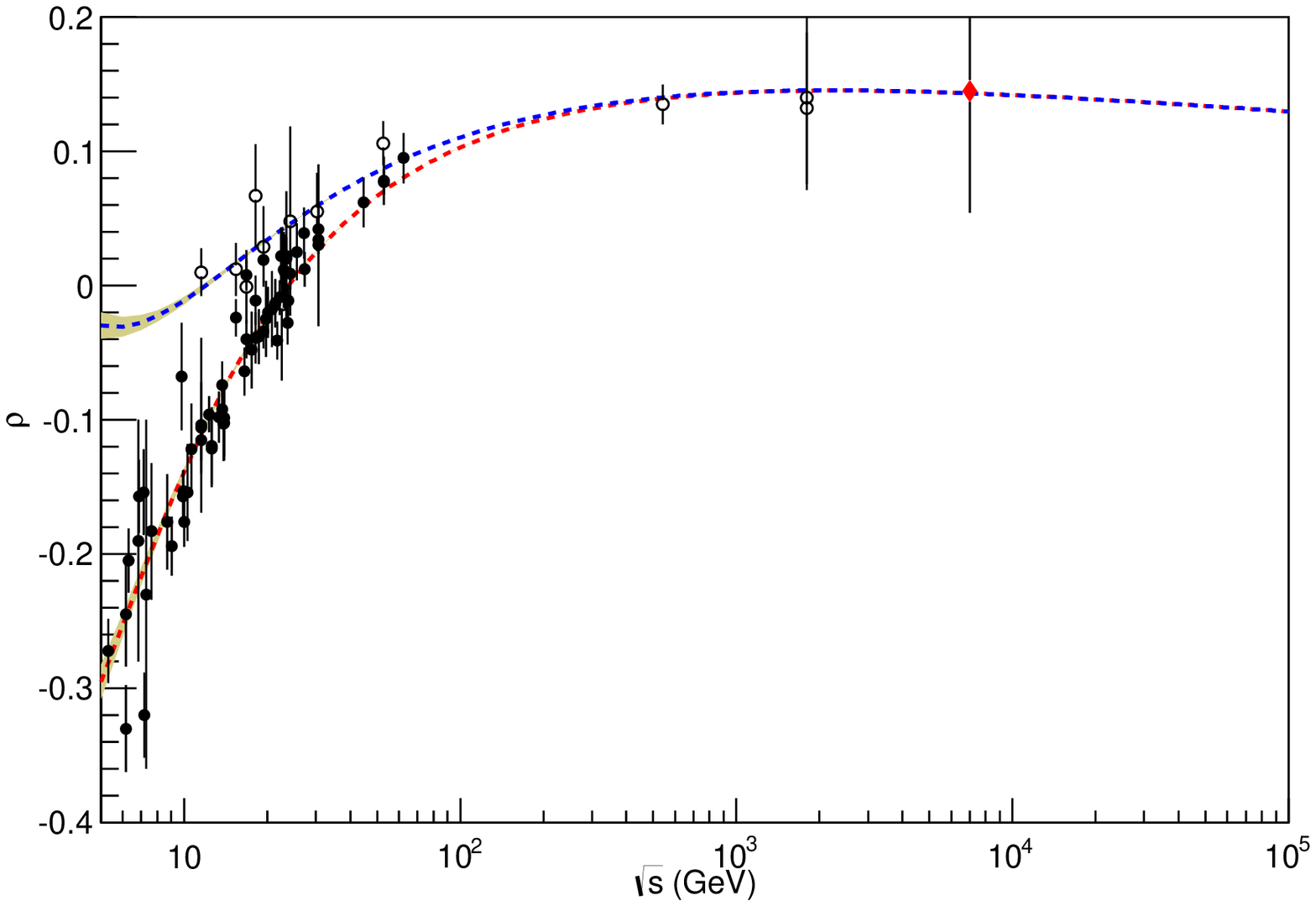,width=15cm,height=8cm}
\caption{Results of the individual fits to $\sigma_{\mathrm{tot}}$ and $\rho$ data 
with the $L\gamma$ model and $s_h = 4m_p^2$ fixed
(table \ref{t5}).
Legend on data as in figure \ref{f1}.}
\label{f5}
\end{figure}

Therefore, with the dataset here considered, the $L\gamma$ model
applies only for $s_h = 4m_p^2$ fixed. 
Note, however, that since the region $s < s_h = 4m_p^2$ (below the 
threshold) constitutes a non-physical region, in this case, model $L\gamma$
is well defined in the whole physical region of scattering states.
The corresponding fit results are displayed in table 
\ref{t5} (second and third columns) and figure \ref{f5}.

\Table{\label{t3}Fit results with the $L2$ model and $s_h$ as a free parameter: 
individual fits to $\sigma_{\mathrm{tot}}$ and
$\rho$ data and extentions to $\sigma_{\mathrm{el}}$ data. Units as in table \ref{t1}.}
\begin{tabular}{c| c| c| c}
\hline
               & $\sigma_{\mathrm{tot}}$ data &    $\rho$ data           &     $\sigma_{\mathrm{el}}$ data \\
\hline
  $a_1$        &   56.2 $\pm$ 7.3    &         56.2             &    117 $\pm$ 66\\
  $b_1$        &  0.588 $\pm$ 0.087  &         0.588            &   1.17 $\pm$ 0.19\\
  $a_2$        &   33.2 $\pm$ 2.3    &         33.2             &      0 (fixed)\\
  $b_2$        &  0.541 $\pm$ 0.016  &         0.541            &      $-$ \\
  $\alpha$     &   37.1 $\pm$ 1.3    &          37.1            &   6.82 $\pm$ 0.12\\
  $\beta$      &  0.312 $\pm$ 0.021  &         0.312            & 0.1200 $\pm$ 0.0064\\
  $\gamma$     &       2 (fixed)     &           2              &      2 (fixed)\\
  $s_h$        &     43 $\pm$ 23     &           43             &    219 $\pm$ 60\\
  $K$          &         $-$         &    48.3 $\pm$ 4.9        &         $-$\\
\hline
  DOF          &         161         &           75             &         103\\
  $\chi^2$/DOF &         0.91        &          1.47            &         1.55\\
  $P(\chi^2)$  &        0.778        &  5.16$\times$10$^{-3}$   &  2.91$\times$10$^{-4}$\\
\hline
Figure:        &    \ref{f3}         &         \ref{f3}         &     \ref{f7}     \\
\hline
  \end{tabular}
\endTable

\Table{\label{t4}Fit results with the $L2$ model and $s_h = 4m_p^2$ fixed: 
individual fits to $\sigma_{\mathrm{tot}}$ and
$\rho$ data and extentions to $\sigma_{\mathrm{el}}$ data. Units as in table \ref{t1}.}
 \begin{tabular}{c|c|c|c}
\hline

               & $\sigma_{\mathrm{tot}}$ data  &    $\rho$ data           &  $\sigma_{\mathrm{el}}$ data \\
\hline
  $a_1$        &    52.1 $\pm$ 2.1    &         52.1             &  270.5 $\pm$ 2.9\\
  $b_1$        &   0.392 $\pm$ 0.019  &         0.392            &  0.480 $\pm$ 0.038\\
  $a_2$        &    33.0 $\pm$ 2.2    &         33.0             &      0 (fixed)\\
  $b_2$        &   0.539 $\pm$ 0.015  &         0.539            &      $-$ \\
  $\alpha$     &   29.75 $\pm$ 0.47   &          29.75           &   3.68 $\pm$ 0.23\\
  $\beta$      &  0.2476 $\pm$ 0.0049 &         0.2476           & 0.0756 $\pm$ 0.0023\\
  $\gamma$     &      2 (fixed)       &           2              &      2 (fixed)\\
  $s_h$        &     3.521 (fixed)    &         3.521            &    3.521 (fixed)\\
  $K$          &          $-$         &     21.9 $\pm$ 4.9       &         $-$     \\
\hline
  DOF          &          162         &           75             &         104\\
  $\chi^2$/DOF &          0.93        &          1.45            &         1.72\\
  $P(\chi^2)$  &         0.718        &  6.71$\times$10$^{-3}$   &  7.77$\times$10$^{-6}$\\
\hline
Figure:        &      \ref{f4}        &              \ref{f4}    &          \ref{f8}      \\
\hline
\end{tabular}
\endTable


\Table{\label{t5}Fit results with the $L\gamma$ model and $s_h = 4m_p^2$ fixed: 
individual fits to $\sigma_{\mathrm{tot}}$ and
$\rho$ data and extentions to $\sigma_{\mathrm{el}}$ data. Units as in table \ref{t1}.}
 \begin{tabular}{c| c| c| c}\hline
               & $\sigma_{\mathrm{tot}}$ data &     $\rho$ data          &  $\sigma_{\mathrm{el}}$ data \\
\hline
  $a_1$        &    60.9 $\pm$ 7.8   &          60.9            &   348.3 $\pm$ 4.9\\
  $b_1$        &   0.530 $\pm$ 0.071 &         0.530            &   0.616 $\pm$ 0.043\\
  $a_2$        &    33.2 $\pm$ 2.3   &          33.2            &      0 (fixed) \\
  $b_2$        &   0.541 $\pm$ 0.016 &         0.541            &      $-$ \\
  $\alpha$     &    34.1 $\pm$ 1.4   &          34.1            &    4.93 $\pm$ 0.13\\
  $\beta$      &   0.102 $\pm$ 0.040 &         0.102            & 0.03122 $\pm$ 0.00081\\
  $\gamma$     &    2.30 $\pm$ 0.14  &          2.30            &     2.30 (fixed)\\
  $s_h$        &     3.521 (fixed)   &         3.521            &    3.521 (fixed)\\
  $K$          &         $-$         &     45.1 $\pm$ 4.9       &          $-$ \\
\hline
  DOF          &          161        &           75             &         104\\
  $\chi^2$/DOF &          0.91       &          1.48            &         1.62\\
  $P(\chi^2)$  &         0.788       &  4.22$\times$10$^{-3}$   &  7.16$\times$10$^{-5}$\\
\hline
Figure:        &     \ref{f5}        &         \ref{f5}         &        \ref{f9}       \\
\hline
 \end{tabular}
\endTable

\subsection{Extensions to elastic cross-section data}
\label{s33}

The connection between the total cross-section and the forward \textit{elastic} amplitude
(optical theorem) led us in \cite{fms2,ms1} to explore the possibility to extend
the same $L\gamma$ model (with $\gamma$ above 2) to fit the elastic cross-section data
(see section 3 in \cite{fms2} for more details). Here we address this extension
taking into account the three models considered.

A noticeable difference between the $\sigma_{\mathrm{tot}}$ and $\sigma_{\mathrm{el}}$ data concerns the
low-energy region, where the evident differences involving $pp$ and $\bar{p}p$ scattering
in the former case are not observed in the latter. For that reason, to extend the parametrization we
consider a degenerate trajectory in the $\sigma_{LE}(s)$ contribution, namely
we fix $a_2$ = 0 in all models.

Concerning the $\sigma_{HE}(s)$ contribution,
since the optical theorem is directly related to unitarity, in applying the same
model for $\sigma_{\mathrm{tot}}$ to fit the $\sigma_{\mathrm{el}}$ data, this principle cannot
be violated, namely for $s \rightarrow \infty$, the ratio $\sigma_{\mathrm{el}}$/$\sigma_{\mathrm{tot}}$
can not go to infinity; moreover, a scenario of an asymptotic transparent disk, namely
$\sigma_{\mathrm{el}}$/$\sigma_{\mathrm{tot}}(s) \rightarrow 0$, is also not expected.
As a consequence, in the case of the $P$ and
$L\gamma$ models, the same values of the exponents in the leading high-energy
contribution ($\epsilon$ and $\gamma$) obtained with the $\sigma_{\mathrm{tot}}$ fit must be considered for the
$\sigma_{\mathrm{el}}$ data reduction.

It is also important to note that, although based on unitarity arguments,
our extension of the parametrization from $\sigma_{\mathrm{tot}}(s)$
to  $\sigma_{\mathrm{el}}(s)$ has here a \textit{strictly empirical character}. In particular,
Regge models have their own results for the elastic differential cross-section
and consequently for the corresponding integrated elastic cross-sections. Therefore, in this section,
the three models will be considered only as empirical parametrizations.

\subsubsection{Fit and results}
\label{s331}

\begin{figure}[pb]
\centering
\epsfig{file=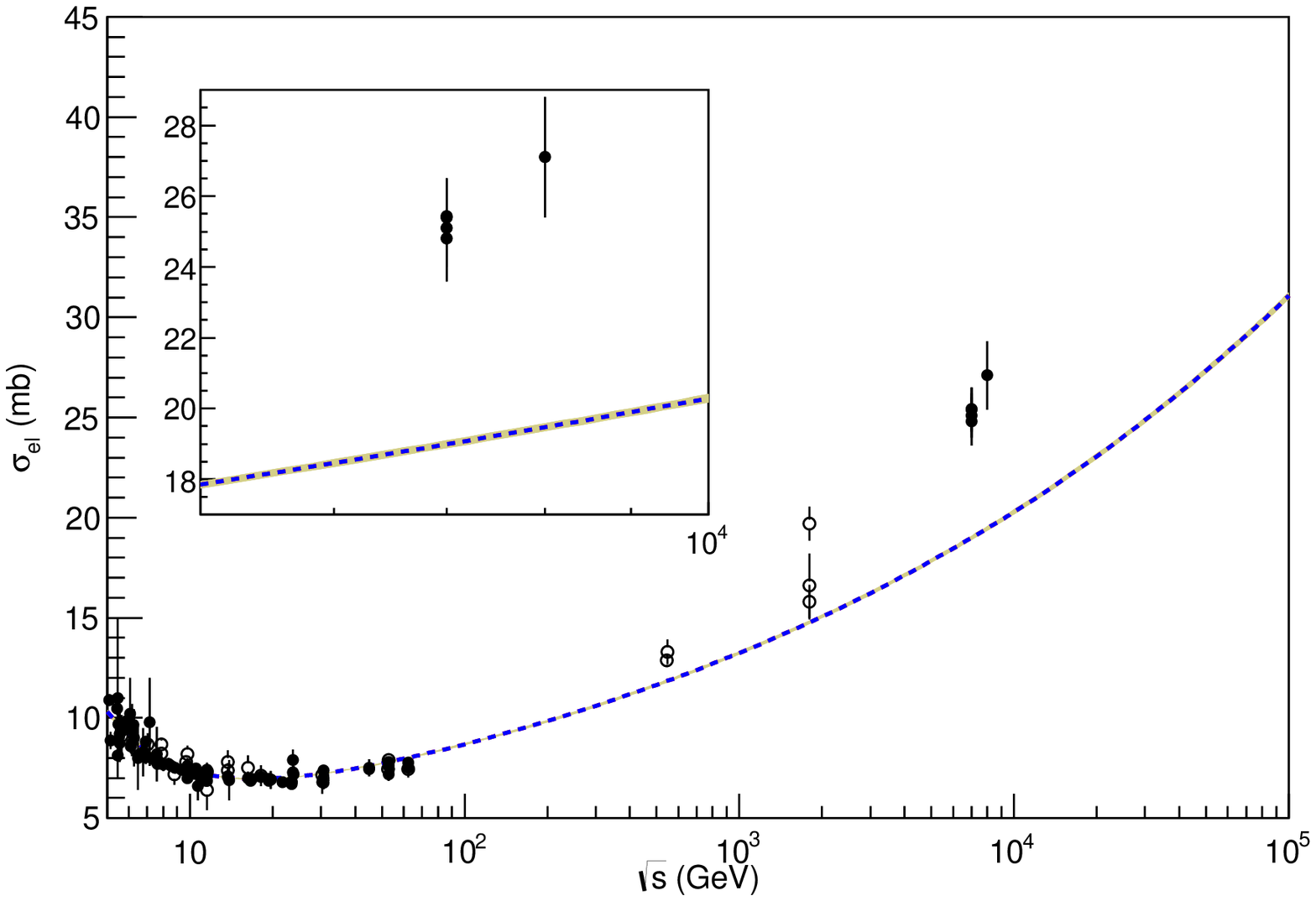,width=15cm,height=8cm}
\epsfig{file=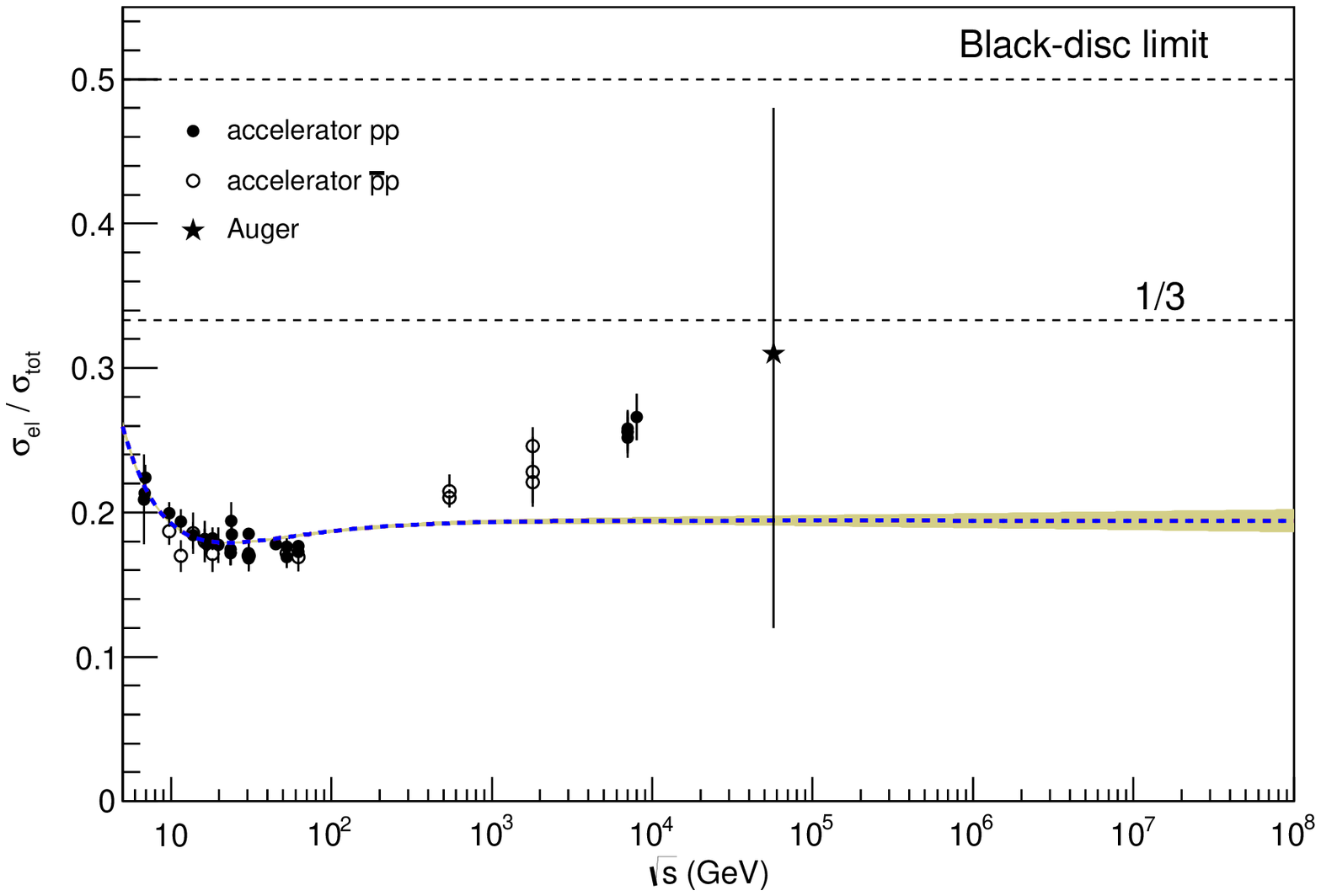,width=15cm,height=8cm}
\caption{Result of the fit to the elastic cross-section data with the $P$ model
and predictions for the ratio between the elastic and total cross-sections (table \ref{t2},
fourth column).}
\label{f6}
\end{figure}

\begin{figure}[pb]
\centering
\epsfig{file=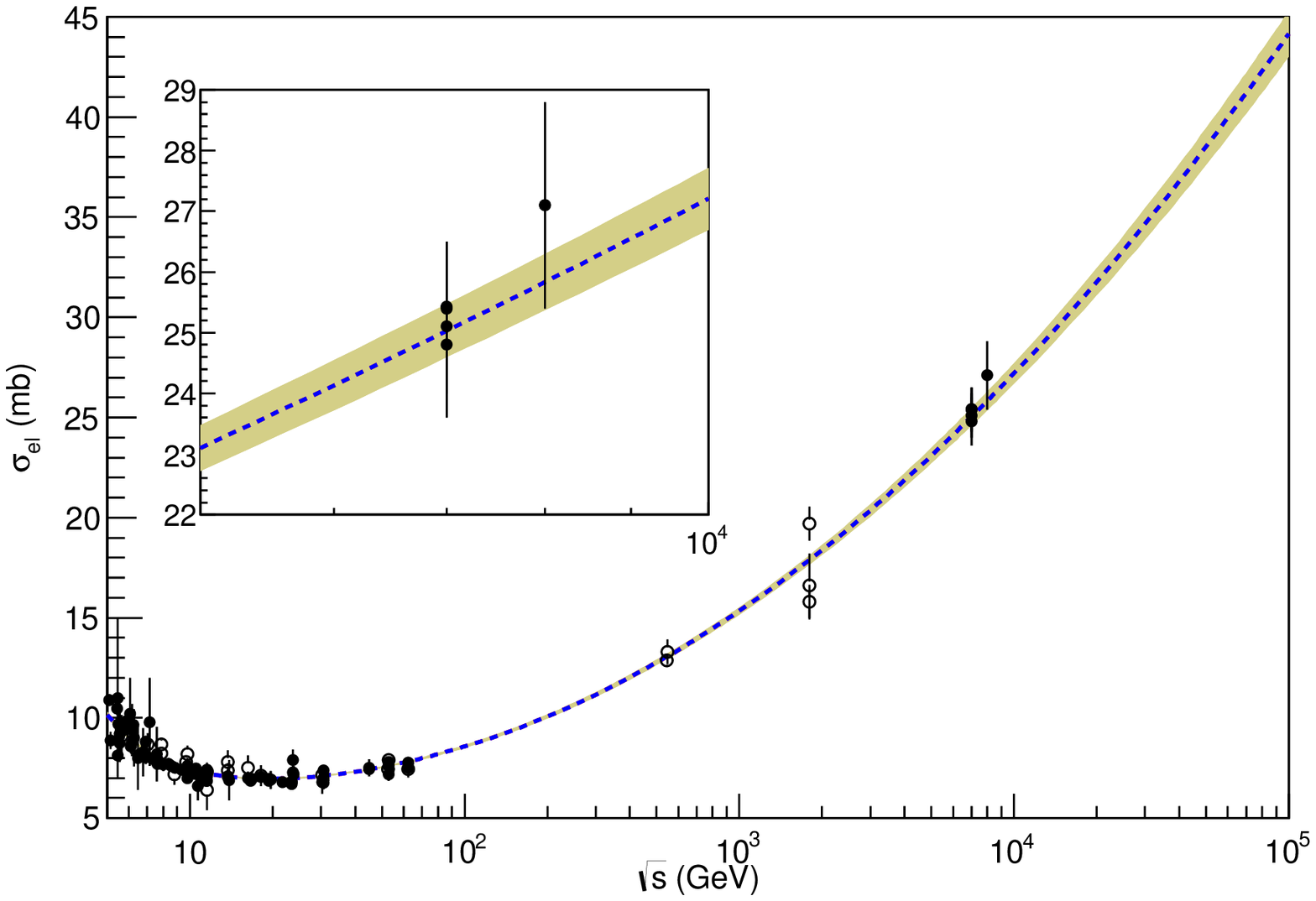,width=15cm,height=8cm}
\epsfig{file=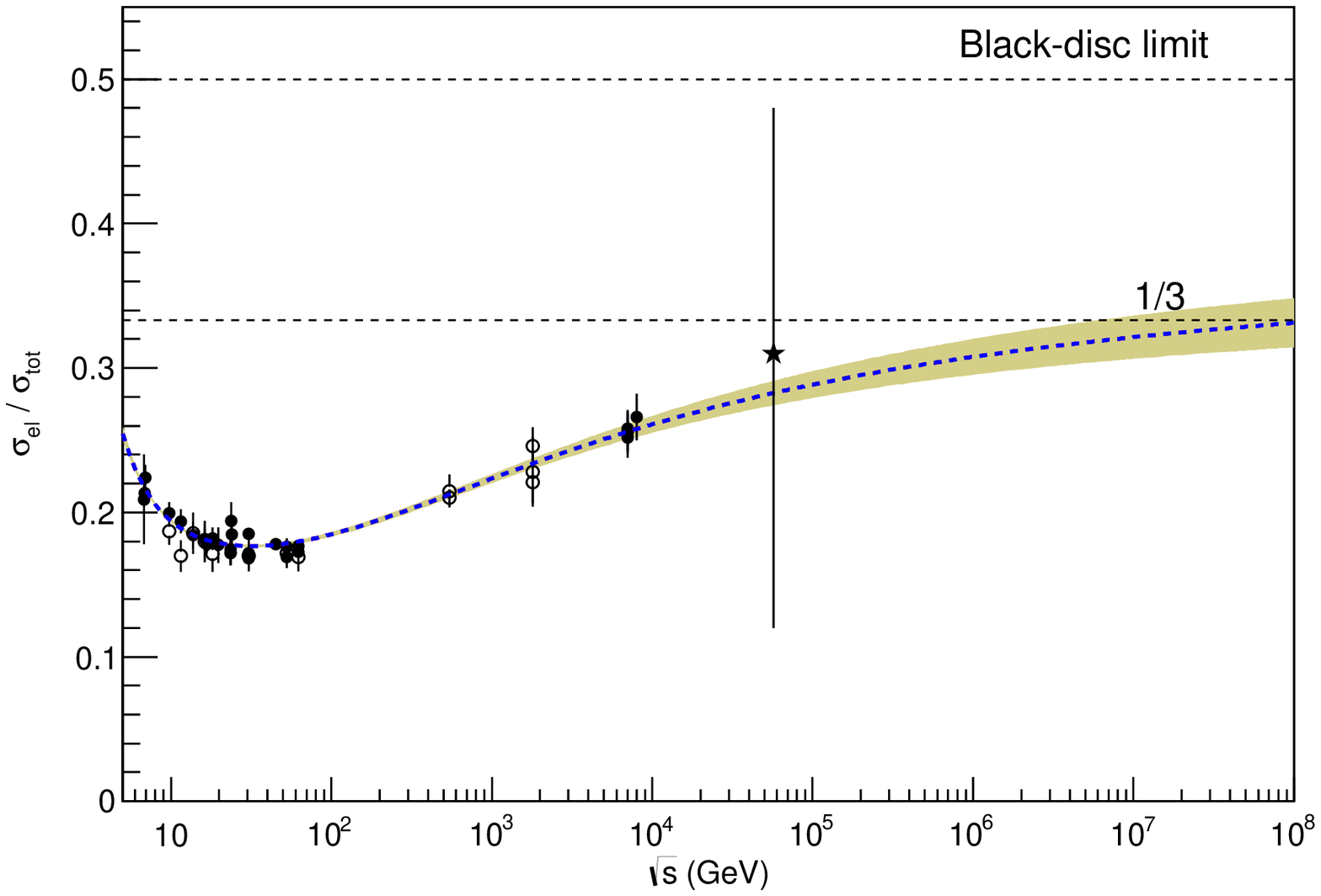,width=15cm,height=8cm}
\caption{Result of the fit to the elastic cross-section data with the $L2$ model in
the case of $s_h$ free and predictions for the ratio between the elastic and total cross-sections
(table \ref{t3}, fourth column).}
\label{f7}
\end{figure}

\begin{figure}[pb]
\centering
\epsfig{file=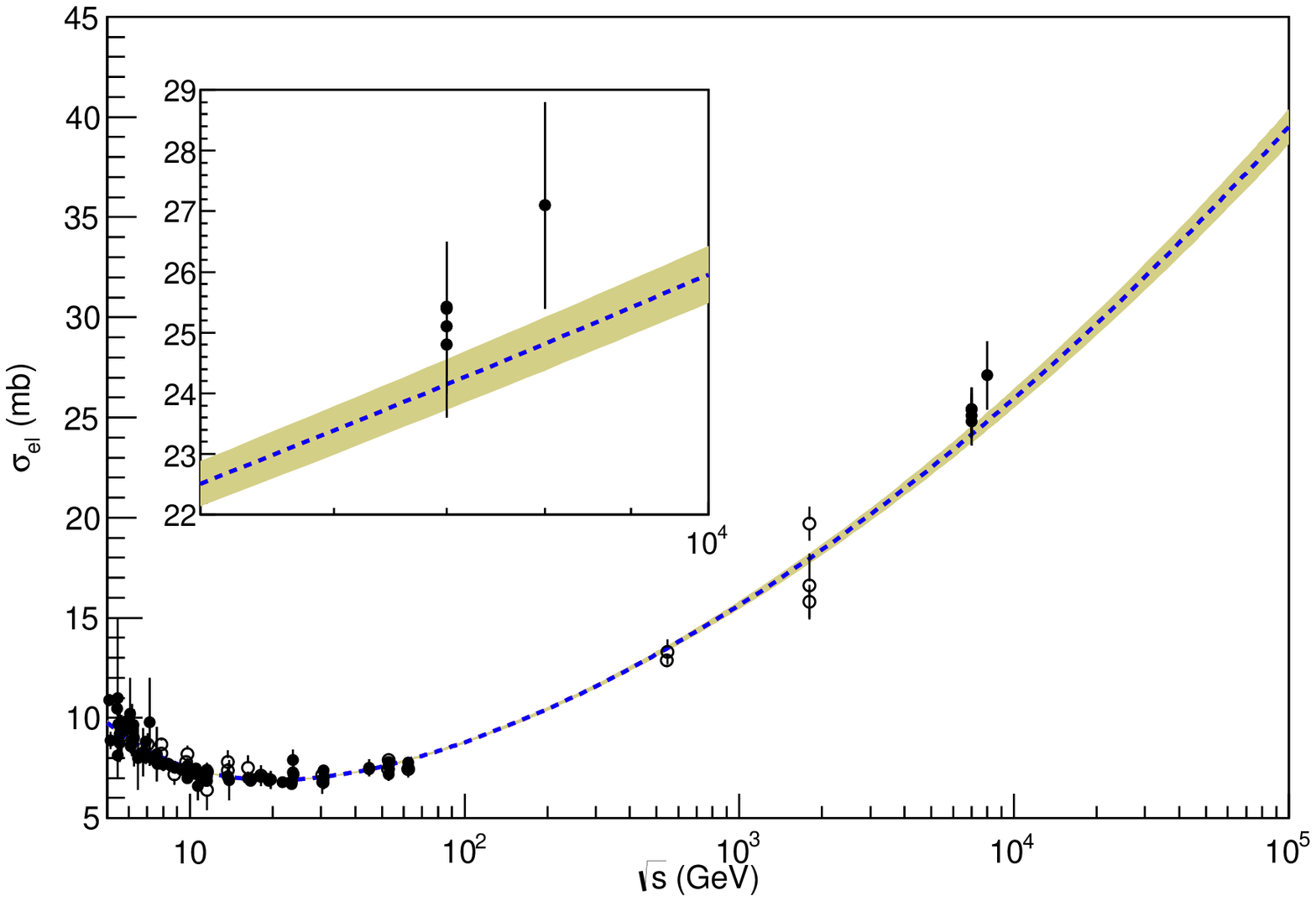,width=15cm,height=8cm}
\epsfig{file=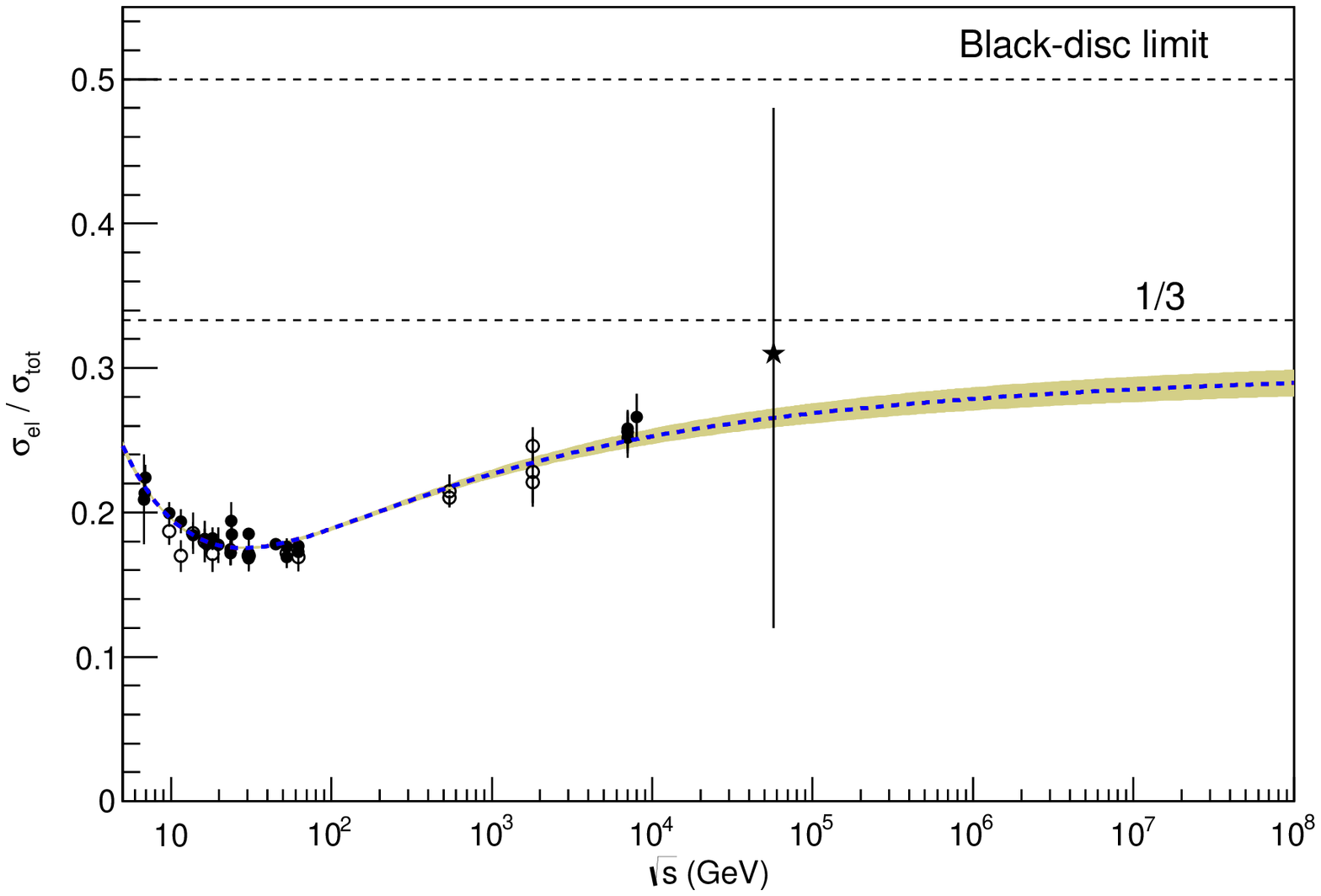,width=15cm,height=8cm}
\caption{Result of the fit to the elastic cross-section data with the $L2$ model in the
case of  $s_h = 4m_p^2$ fixed
and predictions for the ratio between the elastic and total cross-sections
(table \ref{t4}, fourth column).}
\label{f8}
\end{figure}

\begin{figure}[pb]
\centering
\epsfig{file=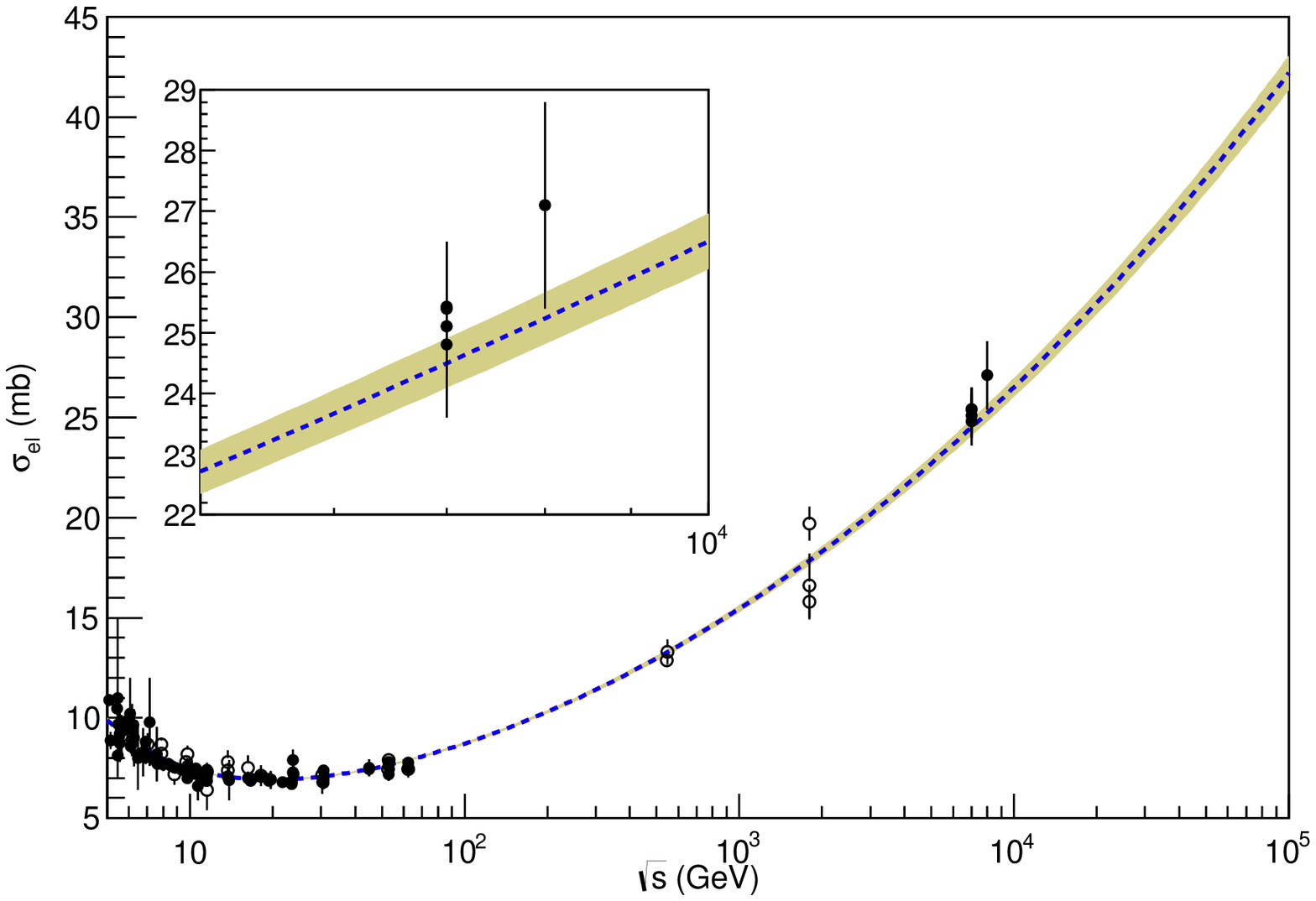,width=15cm,height=8cm}
\epsfig{file=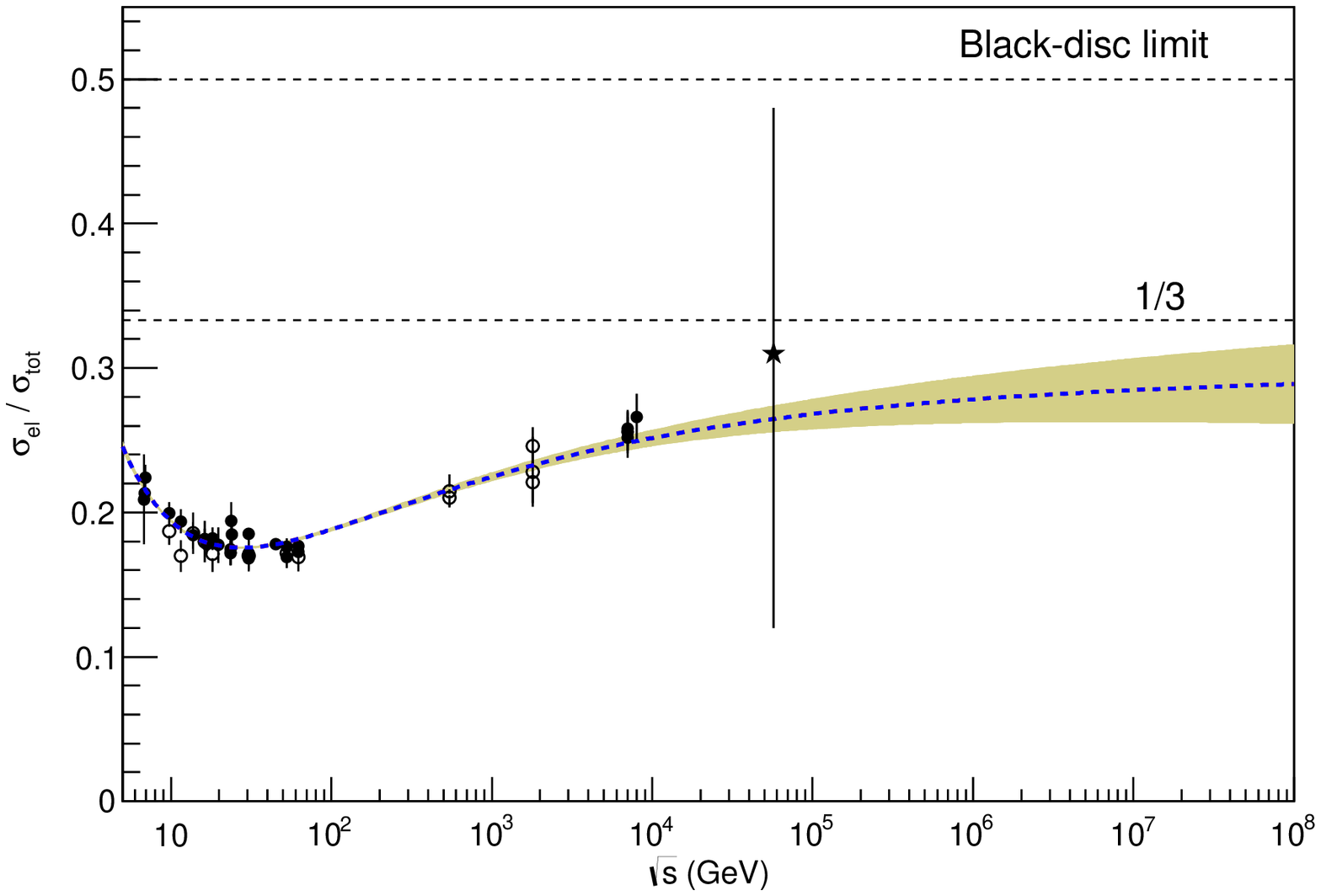,width=15cm,height=8cm}
\caption{Result of the fit to the elastic cross-section data with the $L\gamma$ model
in the case of $s_h = 4m_p^2$ fixed
and prediction for the ratio between the elastic and total cross-sections (table \ref{t5}, fourth column).}
\label{f9}
\end{figure}

Based on the above discussion, in extending the $P$ model to $\sigma_{\mathrm{el}}$
we fix $\epsilon$ = 0.0926 (table \ref{t2}) and in the case of the $L\gamma$
model, $\gamma$ = 2.30 is fixed (table \ref{t5}). As initial values, including the
$L2$ model with the two variants, we use the central values of the parameters obtained 
in the corresponding fits to $\sigma_{\mathrm{tot}}$ data (second columns in tables \ref{t2},
\ref{t3}, \ref{t4} and \ref{t5}). With this procedure, we obtain the fit results
displayed in the fourth columns of tables \ref{t2},
\ref{t3}, \ref{t4} and \ref{t5} and in the upper panels of figures \ref{f6}, \ref{f7}, \ref{f8}, \ref{f9}.
In the lower panels of these figures we show the corresponding predictions for the
ratios between the elastic and total cross-sections.
Using the $s$-channel unitarity, we have also included in these plots the result 
from the estimations of the total cross-section and the inelastic cross-section 
($\sigma_{\mathrm{inel}}$) at 57 TeV by the Auger Collaboration \cite{auger}.

\subsubsection{Asymptotic ratios}
\label{s332}

With the analytic parametrizations (\ref{e4}) - (\ref{e8}),  the value of asymptotic ratio between the elastic and total cross-sections can be 
evaluated. They are related to the parameters $\delta$ ($P$ model) and $\beta$
($L2$ and $L\gamma$ models). Denoting these parameters by the corresponding subscripts
associated with $\sigma_{\mathrm{el}}$ and $\sigma_{\mathrm{tot}}$ fits,
for $s \rightarrow \infty$ we have
\begin{eqnarray} 
\frac{\sigma_{\mathrm{el}}}{\sigma_{\mathrm{tot}}}  \rightarrow \frac{\delta_{\mathrm{el}}}{\delta_{\mathrm{tot}}}
\quad
[P\ \mathrm{model}],
\quad
\frac{\sigma_{\mathrm{el}}}{\sigma_{\mathrm{tot}}}  \rightarrow \frac{\beta_{\mathrm{el}}}{\beta_{\mathrm{tot}}}
\quad
[L2\ \mathrm{and}\ L\gamma\ \mathrm{models}].
\label{e15}
\end{eqnarray} 
From tables \ref{t2} - \ref{t5} we obtain the values displayed in table \ref{t6}.

\Table{\label{t6}Asymptotic results for the ratio $\sigma_{\mathrm{el}}/\sigma_{\mathrm{tot}}$ obtained from the individual fits
to $\sigma_{\mathrm{tot}}$ data and the extensions to $\sigma_{\mathrm{el}}$ data, equation (\ref{e15}).}
\begin{tabular}{c|c}\hline
       Model             &  Asymptotic  Ratio $\sigma_{\mathrm{el}}/\sigma_{\mathrm{tot}}$ \\
\hline
$P$                      &  0.1945 $\pm$ 0.0038\\
$L2$, $s_h$ free         &   0.385 $\pm$ 0.033\\
$L2$, $s_h$ fixed        &   0.305 $\pm$ 0.011\\
$L\gamma$, $s_h$ fixed   &    0.31 $\pm$ 0.12\\
\hline
\end{tabular}
\endTable

\section{Discussion}
\label{s4}

\subsection{Preliminaries}
\label{s41}

Before discussing all results presented in section 3,
it is important to keep in mind four fundamental differences between the
COMPETE analysis and the one developed here.

First, our dataset, restricted to $pp$ and $\bar{p}p$
scattering, constitutes only a sub-set of the data analyzed by the COMPETE
Collaboration. We did not take into account any constraint dictated by other reactions
at low and intermediate energies or a supposed universal behavior.

Second, the COMPETE analysis on $\sigma_{\mathrm{tot}}$ covered the energy region up to 
$\sqrt{s}_{max}$ = 1.8 TeV. This maximal energy is characterized by the disagreement
between the measurement by the CDF Collaboration [70] and the two measurements
by the E710 and E811 Collaborations [71,72]. In fact, as stated in [72], ``the 
confidence level that all 3 measurements are compatible is only 1.6 \%".
On the other hand, our dataset includes all the
high-precision TOTEM measurements at 7 TeV (4 points) and 8 TeV (1 point).

Third, the COMPETE has employed a detailed procedure of ranking models,
including seven distinct statistical indicators. In our case, only the $\chi^2$/DOF and
$P(\chi^2)$ have been used to check the statistical consistency of the data reductions
in a reasonable way.

Fourth, taking into account several classes of analytic parametrizations and constraints,
the COMPETE Collaboration investigated 256 variants, selecting 24 possible models under the criteria
of $\chi^2$/DOF $\leq$ 1.0 and non-negative pomeron contribution at all energies \cite{compete1}.
Among those models they have eventually selected their highest rank parametrization.
Here, we have restricted our analysis to only three analytic models
(and two variants in one case). Nonetheless, the novel aspect concerns the use of the
parametrization introduced by Amaldi \textit{et al.}, which, unfortunately,
did not take part in the COMPETE analysis. 

Bearing in mind the above differences, let us now discuss all the results presented 
in section \ref{s3}, including reference
to some results by the COMPETE Collaboration. We shall discuss separately the data reductions for the total 
cross-section and $\rho$ parameter (section \ref{s42}) and the elastic cross-section (section \ref{s43}).

\subsection{Results for the total cross-section and $\rho$}
\label{s42}

The main point here is to confront the results obtained through model $L2$ 
with those provided by models $P$ and $L\gamma$.
To this end, we divide the discussion as follows.

\subsubsection{$L2$ model \textit{versus} $P$ model}
\label{s421}

Let us start with the COMPETE results displayed in figure 1, table 1, related
to the $P$ model and the $L2$ model.
Although both extrapolations led to good agreement with the
TOTEM results, within the uncertainties, the compatibility in the case of the $L2$ model, especially at
8 TeV, is indeed striking. Even the statistical result with our dataset is quite
good: $\chi^2$/DOF = 1.01 (table \ref{t1}). 
We note, however, that in the COMPETE result, $s_h \sim $ 34 GeV$^2$ which means that
with the assumed energy cutoff at $s_{min}$ = 25 GeV$^2$, the leading $L2$ pomeron
contribution decreases as the energy increases in the region from $s_{min}$
up to $s_h$ (as already discussed in \cite{ms1}, section 4.2, figure 7).

With regard to the $P$ model predictions, if compared with the above $L2$ model result,
the extrapolation, although in agreement with the TOTEM data within the uncertainties,
overestimates the \textit{central} values of the TOTEM results at 7 and 8 TeV (figure \ref{f1}).
We recall that in \cite{compete2}, the COMPETE
Collaboration does not consider this model and in \cite{compete1}, the authors conclude
that it ``fails to reproduce jointly the total cross section and the $\rho$ parameter
for $\sqrt{s} \geq$ 5 GeV''.

Let us turn now  to our fit results with the dataset here considered: table \ref{t2} and 
figure \ref{f2} ($P$ model)
and tables \ref{t3}, \ref{t4}, figures \ref{f3}, \ref{f4} ($L2$ model in the
cases of $s_h$ free and fixed, respectively). 
Here, the situation is rather different, since in both cases, the fit results are essentially 
equivalent on statistical grounds: $P(\chi^2) \approx$ 0.8 with the $P$ model and $L2$ model
in the case of $s_h$ free and $P(\chi^2) \approx$ 0.7 with the 
$L2$ model and $s_h$ fixed. Moreover, in the LHC region, the results
of the $P$ model and the $L2$ model ($s_h$ free) are practically
identical
(figures \ref{f2} and \ref{f3}):
the upper band of the fit uncertainty region includes the central values of the
TOTEM data at 7 TeV and the upper extreme of the band reaches the central value
of the datum at 8 TeV. The same equivalence can be verified in the corresponding
results for $\rho(s)$ and the descriptions of the data are quite good. 
However, we note that, as in the COMPETE result, the $L2$ model with $s_h$ free yields 
$s_h \sim $ 43 GeV$^2$ (table \ref{t3}) and therefore a decreasing pomeron contribution between
$s_{min}$ = 25 GeV$^2$ and $s_h$. That obviously is not the case with the $P$
model, since it brings enclosed the rise of the leading high-energy contribution at all energies.

On the other hand, \textit{compared with the above results}, the $L2$ model 
in the case of $s_h = 4m_p^2 \sim $ 3.5 GeV$^2$ fixed 
(far below the cutoff) partially underestimates the TOTEM data (including the central
value of the $\rho$ estimation at 7 TeV), as shown in figure \ref{f4}
and discussed below in section \ref{s422}.

We note that, with the $P$ model, the value of the soft pomeron intercept obtained here reads
$\alpha_{\mathcal{P}}(0) = 1 + \epsilon = 1.0926 \pm 0.0016$. 
This result can be compared with those reported in some previous analyses, as
shown in table \ref{t7} ($\sqrt{s}_{max}$ = 1.8 TeV). We see that, within the uncertainties, our result is in 
agreement with those obtained by Cudell, Kang and Kim,  also by Cudell \textit{et al.} and 
with the
extrema bounds inferred by Luna and Menon; our result lies slightly below the COMPETE
result (table \ref{t1}) and above the historical value by Donnachie and Landshoff
(degenerate trajectories).

\Table{\label{t7}Values of the soft pomeron intercept ($\alpha_{\mathcal{P}}(0) = 1 + \epsilon$) from
some previous analyses and from this work.}
\begin{tabular}{c|c}\hline
                           &  $\alpha_{\mathcal{P}}(0)$ \\
\hline
Donnachie-Landshoff (1992)   \cite{dl3}         & 1.0808 \\
Cudell, Kang and Kim (1997) \cite{ckk}          &  1.096$_{-0.009}^{+0.012}$ \\
Cudell \textit{et al.} (2000) \cite{cudell2000} &  1.093 $\pm$ 0.003         \\
COMPETE Collaboration (2002) \cite{compete1}    &  1.0959 $\pm$ 0.0021     \\
Luna and Menon (2003)  \cite{lm}                &  1.085 - 1.104      \\
This work                                       & 1.0926 $\pm$ 0.0016 \\
\hline
\end{tabular}
\endTable

\subsubsection{$L2$ model \textit{versus} $L\gamma$ model}
\label{s422}

With the $L\gamma$ model we obtain
a convergent fit solution only in the case of $s_h$ fixed. Therefore, our
comparative discussion with the $L2$ model will be restricted here to
this variant (individual and global fits).

As regards individual fits, from tables \ref{t4} and \ref{t5}, with the
$L2$ model $P(\chi^2) \approx$ 0.72 and in the case of the $L\gamma$
model, $P(\chi^2) \approx$ 0.79, indicating, therefore, a practical equivalence on
statistical grounds. On the other hand, noticeable differences can be identified
in the LHC region, as shown in figures \ref{f4} and \ref{f5}. In the case
of the $L2$ model (figure \ref{f4}), the fit uncertainty region shows agreement with
only the lower error bar of the TOTEM results at 7 TeV and the upper extreme of the band does not
reach the central value of the datum at 8 TeV.
That, however, is not the case with the $L\gamma$ model (figure \ref{f5}), since the uncertainty
region includes all the central values of the data at 7 TeV and the upper band
reaches the central value at 8 TeV.

With these models and variant, we have also developed simultaneous fits to 
$\sigma_{\mathrm{tot}}$ and $\rho$ data, which are presented in appendix B:
table \ref{t8} (second and fourth columns) and figures \ref{f11} and \ref{f12}.
In both cases, we obtain a smaller integrated probability, $P(\chi^2) \approx$ 0.2,
which is a consequence of the inclusion of the $\rho$ data. From the figures, the results
at the LHC region  are practically identical to those obtained in the individual fits.

With the $L\gamma$ model, in going from individual to global fits, we notice a slightly
decrease in the $\gamma$ parameter, from 2.30 (table \ref{t5}) to 2.23 (table \ref{t8}).
This difference is also due to the inclusion of the $\rho$ data, which constraint
the rise of the total cross section (see our discussion in \ref{saa2}).

\subsubsection{Conclusions on the fit results}
\label{s423}

Based on the results presented here and on the above discussion, we are led to the 
conclusions that follow.

\begin{description}

\item{1.}
The fit results with the $P$ model, the $L2$ model (in the case of $s_h$ free) and the
$L\gamma$ model ($s_h$ fixed) are all consistent within their uncertainties, leading to
equivalent descriptions of the experimental data. That means three different possible
scenarios for the rise of the total cross-section at the highest energies.

\item{2.}
In all cases above, the fit results are not in plenty agreement with the TOTEM
datum at 8 TeV: within the uncertainties, the data reductions partially underestimate
this high-precision measurement.

\item{3.}
Compared with the results mentioned above,
the $L2$ model with the variant $s_h$ fixed leads to fit results less
consistent with the experimental data. In the case of $s_h$ free, however, the leading
pomeron contribution decreases as the energy increases below $\sqrt{s_h} \approx$ 7 GeV
(table \ref{t3}).

\item{4.}
As regards models $P$ and $L\gamma$, in addition to their
efficiency in the description of the experimental data (except, perhaps, at 8 TeV),
both bring enclosed an increasing pomeron contribution for all values
of the energy (above the threshold $4m_p^2$ in the last case).

\item{5.}
For $s_h = 4m_p^2$ fixed, model $L\gamma$ is well defined in the
whole physical region of scattering states and the fit results
are more consistent with the $\sigma_{tot}$ data at 7 and 8 TeV
than model $L2$ (figures \ref{f4} and \ref{f5} and also 
\ref{f11} and \ref{f12}).

\end{description}

\subsection{Results for the elastic cross-section and asymptotia}
\label{s43}

\subsubsection{Conclusions on the fit results.}
\label{s431}

Concerning the $P$ model here considered, despite the efficient descriptions of the
$\sigma_{\mathrm{tot}}$ and $\rho$ data, it is certainly not adequate to be extended to fit the
$\sigma_{\mathrm{el}}$ data with $\epsilon$ fixed, as shown in table \ref{t2} (fourth column) and figure \ref{f6}.
We have displayed these results only as a complementary information and we shall not refer
to them in what follows.

In the cases of the $L2$ and $L\gamma$ models, despite the rather
small integrated probabilities (tables \ref{t3}, \ref{t4} and \ref{t5},
fourth columns), the description of the experimental data
below the LHC region is quite good in all cases 
(figures \ref{f7}, \ref{f8} and \ref{f9}). The differences (and drawbacks) concern
the TOTEM results at 7 TeV (four points) and 8 TeV (one point), as shortly discussed
in what follows.

With the $L2$ model and $s_h$ free (fig. \ref{f7}), the TOTEM data at
7 TeV are quite well described within the uncertainties and the fit uncertainty
region includes part of the lower error bar at 8 TeV. The corresponding prediction
for the ratio $\sigma_{\mathrm{el}}$/$\sigma_{\mathrm{tot}}$ is also in good 
agreement with the TOTEM data at 7 and 8 TeV, within the uncertainties.

In the case of $s_h$ fixed (fig. \ref{f8}) the uncertainty region of the fit 
with the $L2$ model is consistent
with the lower error bar at 7 TeV, but does not reach the lower error bar at 8 TeV.
Analogous results are obtained with the $L\gamma$ model (fig. \ref{f9}),
except that, in this case, the fit uncertainty region reaches the lower error bar
at 8 TeV.

We have also extended to $\sigma_{\mathrm{el}}$  the results obtained in global fits to 
$\sigma_{\mathrm{tot}}$ and $\rho$ data with the $L2$ and $L\gamma$ models ($s_h$
fixed), presented in Appendix B. From table \ref{t8} (third and fifth columns)
and figures \ref{f11} and \ref{f12}, we are led to the same conclusions outlined above.

\begin{figure}[pb]
\centering
\epsfig{file=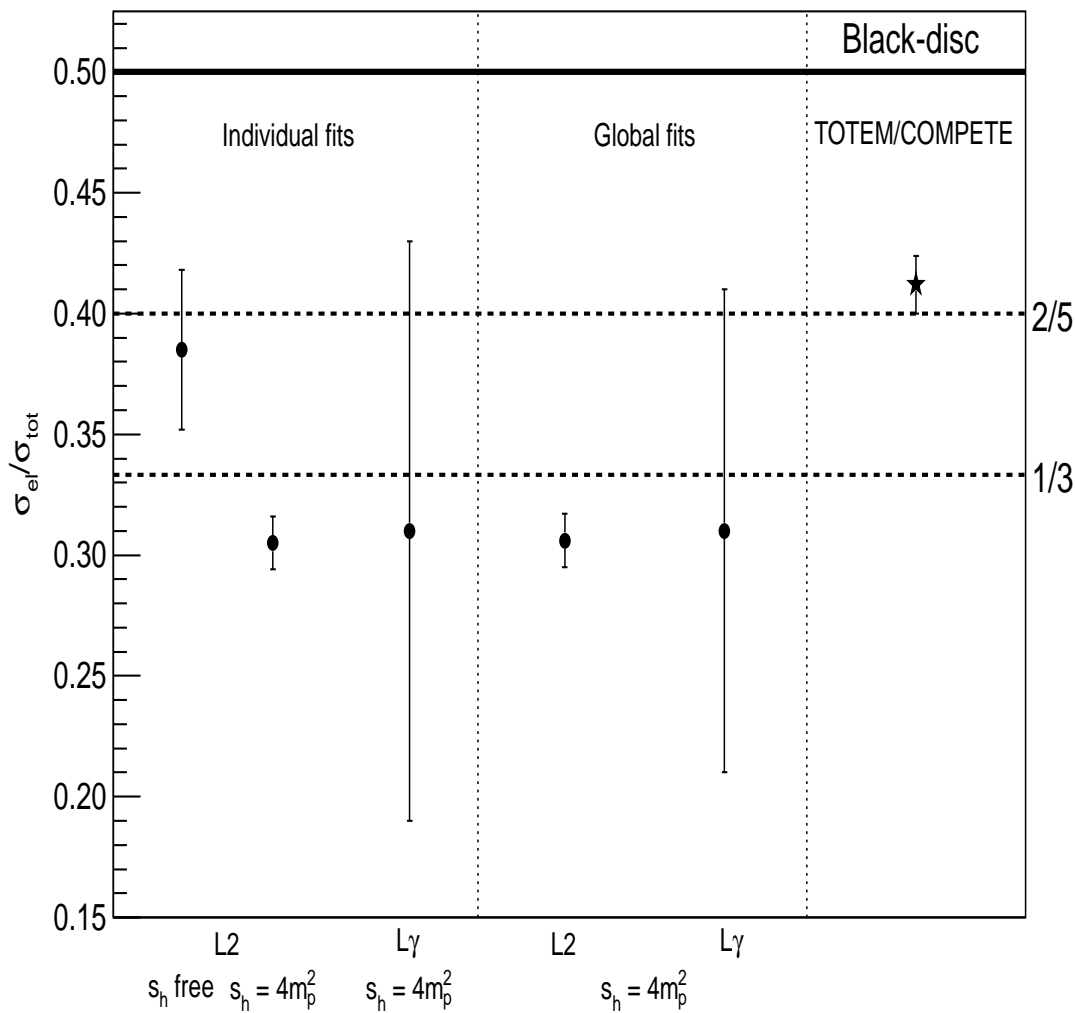,width=15cm,height=15cm}
\caption{Asymptotic ratios between the elastic and total cross-sections obtained in this analysis
and combining the TOTEM and COMPETE results (tables \ref{t6}, \ref{t9} and equation (\ref{e16})).}
\label{f10}
\end{figure}

\subsubsection{Asymptotia.}
\label{s432}

The asymptotic ratios between the elastic and total cross-sections obtained here
are displayed in table \ref{t6} (individual fits) and table \ref{t9} (global
fits). Neglecting reference to the $P$ model result, all ratios 
(individual and global fits) do not exceed 0.430,
\textit{within the uncertainties} (namely the upper uncertainty bounds),
indicating, therefore, asymptotic values below 0.5, the black-disk limit.
These values within their uncertainties are schematically summarized in figure
\ref{f10} (including one point to be discussed below).

It may be interesting to note that the TOTEM Collaboration usually displays in their
figures an analytical (empirical) fit to the elastic cross section data \cite{totem3,totem4}
\begin{eqnarray} 
\sigma_{\mathrm{el}}(s) = 11.4 - 1.52 \ln{s} + 0.130 \ln^2{s},
\nonumber
\end{eqnarray} 
which, with our notation in section \ref{s332}, equation (\ref{e15}), indicates
\begin{eqnarray} 
\beta_{\mathrm{el}}^{\mathrm{TOTEM}} = 0.130\ \mathrm{mb}.
\nonumber
\end{eqnarray} 
Now, if we use the 2002 COMPETE result for the total cross section, from table \ref{t1},
we have
\begin{eqnarray} 
\beta_{\mathrm{tot}}^{\mathrm{COMPETE}} = 0.3152 \pm 0.0095\ \mathrm{mb},
\nonumber
\end{eqnarray} 
and we can infer (for $s \rightarrow \infty$)
\begin{eqnarray} 
\frac{\sigma_{\mathrm{el}}}{\sigma_{\mathrm{tot}}} \rightarrow
\frac{\beta_{\mathrm{el}}^{\mathrm{TOTEM}}}{\beta_{\mathrm{tot}}^{\mathrm{COMPETE}}} = 0.412 \pm 0.012.
\label{e16}
\end{eqnarray} 
This point, also displayed in figure \ref{f10} (right), corroborates, within all uncertainties,  
the upper bound 0.430 mentioned above.

Based on figure \ref{f10} and tables \ref{t6} and \ref{t9}, we are led to the
following conclusions on the asymptotic ratio $\sigma_{\mathrm{el}}/\sigma_{\mathrm{tot}}$.

\begin{description}

\item{1.}
The results from the $L2$  and  $L\gamma$ models with $s_h$ fixed (individual and global fits)
are all consistent within the uncertainties and predict a ratio around 0.3 (fairly below a rational
limit 1/3).

\item{2.}
Within the uncertainties, the results of model $L\gamma$ (individual and global fits)
are consistent with the rational limit 1/3, as obtained in previous analyses with this
model \cite{fms2,ms1}.

\item{3.}
The result from the $L2$ model with $s_h$ free (individual fit) and the estimated ratio
from the TOTEM and COMPETE parametrizations are almost consistent within the uncertainties,
indicating a ratio above 1/3, possibly, around 0.4 = 2/5.

\item{4.}
All the results here obtained are consistent with an asymptotic ratio below the black-disk limit
and, in terms of rational values, it seems plausible to estimate
\begin{eqnarray} 
\frac{1}{3} \lesssim \frac{\sigma_{\mathrm{el}}}{\sigma_{\mathrm{tot}}}  \lesssim \frac{2}{5}
\quad
\mathrm{as\ }
s \rightarrow \infty.
\nonumber
\end{eqnarray}

\end{description}

As previously conjectured by Grau \textit{et al.} \cite{grau} and discussed
in \cite{fms2,ms1}, this
result can be interpreted as a combination of the soft scattering states
(elastic and diffractive), giving rise to the
black-disk limit. In a formal context, it points towards 
a saturation of the Pumplim bound \cite{pump1,pump2,suk},
\begin{eqnarray} 
\frac{\sigma_{\mathrm{el}}}{\sigma_{\mathrm{tot}}}  + \frac{\sigma_{\mathrm{diff}}}{\sigma_{\mathrm{tot}}}
\leq \frac{1}{2},
\nonumber
\end{eqnarray} 
where $\sigma_{\mathrm{diff}}$ is the cross-section associated with the inelastic
dissociation processes. In this context, our above estimation indicates

\begin{eqnarray} 
\frac{1}{10} \lesssim \frac{\sigma_{\mathrm{diff}}}{\sigma_{\mathrm{tot}}}  \lesssim \frac{1}{6}
\quad
\mathrm{as\ }
s \rightarrow \infty.
\nonumber
\end{eqnarray} 

At last, we recall that these results and arguments contrast with the asymptotic 
black-disk scenario predicted 
in the model-dependent 
amplitude analysis
by Block and Halzen \cite{bhbd}.

\section{Summary, conclusions and final remarks}
\label{s5}

We have presented a comparative study on three analytic parametrizations for the
hadronic total cross-section, distinguished by their high-energy leading
(pomeron) contributions. Including the non-degenerate Reggeon terms for the low
and intermediated energy region, the parametrizations have been denoted as 
models $P$,
$L2$ and $L\gamma$. The analytic connection with the $\rho$
parameter has been obtained by means of singly subtracted derivative dispersion relations (DDR),
with the corresponding subtraction constant as a free fit parameter.

As regards the practical equivalence between integral dispersion relations and DDR 
(without the high-energy approximation),
we have discussed in appendix A the fundamental role of the subtraction constant.
We have also observed that in the COMPETE analysis, reference is made on the use of
DDR, but without information on the subtraction constant \cite{compete1,compete2}.

Our dataset comprised only $pp$ and $\bar{p}p$ scattering, but covering the energy-region
from 5 GeV up to 8 TeV. Individual and global fits to $\sigma_{\mathrm{tot}}$ and
$\rho$ data have been addressed and also extensions to $\sigma_{\mathrm{el}}$ data with the corresponding
extraction of the asymptotic ratios between $\sigma_{\mathrm{el}}$ and $\sigma_{\mathrm{tot}}$.
One important and, presently, yet novel aspect of our analysis is the inclusion in the dataset
of all the experimental information presently available at 7 TeV and 8 TeV \cite{totem1,totem2,totem3,totem4}.

Based on the results and discussions presented here, we are led to the following four main conclusions:

\begin{description}

\item{1.} 
The data reductions with models $L2$ and $L\gamma$  are strongly dependent on the 
high-energy-scale factor $s_h$.

\item{2.} 
The fit results to $\sigma_{\mathrm{tot}}$ and $\rho$ data with models $P$, the $L2$ 
($s_h$ free) and $L\gamma$ ($s_h$ fixed and $\gamma$ above 2) are all consistent
within their uncertainties and with
the experimental data up to 7 TeV. However, the data reductions partially underestimate the 
high-precision TOTEM measurement at 8 TeV.

\item{3.} 
Once compared with the above results, model $L2$  with $s_h$ fixed is less consistent with the data 
and in the case of $s_h$ free, the leading high-energy pomeron contribution decreases as the energy
increases below $\sqrt{s_h} \approx$  7 GeV.

\item{4.}
With models $L2$ and $L\gamma$ (degenerate trajectories), the extensions of the parametrizations 
to fit the  $\sigma_{\mathrm{el}}$ data led to asymptotic ratios between  $\sigma_{\mathrm{el}}$ and $\sigma_{\mathrm{tot}}$
below the black-disk limit, within the uncertainties.
The results favor asymptotic
rational limits in the interval 1/3 - 2/5
and points towards a saturation of the
Pumplin bound.

\end{description}

It is important to emphasize two contrasting physical pictures present in our results and
including the 2002 COMPETE result.
For $\sigma_{\mathrm{tot}}$, we have, on the one hand, model $P$  and models
$L2$ and $L\gamma$ with $s_h = 4m_p^2$ implying in an increasing monotonic pomeron contribution
for all values of the energy (above the threshold in the last two cases);
on the other hand, model $L2$ with $s_h$ free predicting a decreasing pomeron contribution
as the energy increases below $s_h$. Therefore, in the energy region investigated (above the energy cutoff),
two contrasting physical pictures emerge, involving only one model/variant (including
the COMPETE result) and all the other three cases. That calls into question whether this decreasing
effect in the pomeron contribution has a fundamental theoretical/phenomenological justification
or is a consequence of the data reduction. That seems a key issue
because, as we have shown (also in \cite{ms1}), this effect is directly related with the striking
agreement of the COMPETE extrapolation with the high-precision TOTEM measurements at 7 TeV and 8 TeV.

In our introduction, we have quoted two questions put in our first work on this subject
\cite{fms1}.
Based on the results presented here and in our previous analyses \cite{fms1,fms2,ms1},
we understand that we have collected enough material to improve the answers to these questions
without over interpretations:
(1)
model $L2$ does not represent a unique solution describing the asymptotic
rise of the total cross-section; 
(2)
the available data can as well be statistically described by model $P$ with 
$\epsilon \approx$ 0.093 (table \ref{t2}) and by  model $L\gamma$
with $\gamma \approx$ 2.3 (tables \ref{t5} and \ref{t8}).

Nonetheless, just answering the two questions, our final conclusion, as already
stressed in previous works, is that the
rise of the hadronic total cross-section at the highest energies still constitutes an open problem,
demanding, therefore, further and detailed investigation. 
In this respect, we understand that
an updated analysis, \textit{including all the experimental data currently available} and
along detailed procedures as those developed by the COMPETE Collaboration (more than ten years ago) 
can certainly provide
new and updated insights on the subject, even before the future experimental data at 13-14 TeV.
Moreover, updated \textit{model-independent} analyses on the rise of the ratio between the elastic and total cross
section (as, for example, that developed in \cite{fm1,fm2}), may also shed some light on the subject of asymptotia.

\vspace{0.3cm}

%\section*{Acknowledgments}

\ack

We wish to thank Ya. I. Azimov for fruitfull correspondence, 
C. Dobrigkeit and D.A. Fagundes for useful discussions and suggestions.
We are also thankful to two anonymous referees for valuable
comments, suggestions and discussions.
Research supported by FAPESP 
(Contracts Nos. 11/15016-4, 09/50180-0).

\appendix
\section{Derivative dispersion relations and the subtraction constant}
\label{saa}

In this appendix, we treat the analytical connection between $\sigma_{\mathrm{tot}}(s)$ and $\rho(s)$
using singly-subtracted derivative dispersion relations (DDR). Specifically, the point is to deduce
equations (\ref{e9}) - (\ref{e14}) for $\rho(s)$ from the analytical parametrizations for
$\sigma_{\mathrm{tot}}(s)$, equations  (\ref{e4}) - (\ref{e8}). Although 
several aspects of this connection had already been
discussed in our previous works \cite{fms1,fms2}, some points associated with the practical use of the
derivative dispersion relations and the role of the subtraction constant deserve to be stressed.
In what follows, after reviewing the main formulas related to integral and derivative
dispersion relations (section A.1), we present some critical comments on the practical use of the derivative
relations (section A.2).

\subsection{Analytic results}
\label{saa1}

For $pp$ and $\overline{p}p$ scattering, analyticity and crossing symmetry
allow us to connect $\sigma_{\mathrm{tot}}(s)$ and $\rho(s)$ through the
formulas \cite{pred}
\begin{eqnarray}
\rho ^{pp}(s) \sigma_{\mathrm{tot}}^{pp} (s) =
\frac{\textrm{Re}\,F_{+}}{s}  + \frac{\textrm{Re}\,F_{-}}{s},
\label{ea1}
\end{eqnarray}

\begin{eqnarray}
\rho ^{\bar{p}p}(s) \sigma_{\mathrm{tot}}^{\bar{p}p} (s) =
\frac{\textrm{Re}\,F_{+}}{s}  - \frac{\textrm{Re}\,F_{-}}{s},
\label{ea2}
\end{eqnarray}
where the even ($+$) and odd ($-$) amplitudes are related to the $pp$ and $\overline{p}p$
amplitudes by

\begin{eqnarray}  
F_{\pm}(s) = \frac{F^{pp} \pm 
F^{\bar{p}p}}{2}.
\label{ea3}
\end{eqnarray} 

Dispersion relations have been first deduced in the  \textit{integral} form and
in the case of the forward direction the standard once-subtracted
\textit{integral dispersion
relations} (IDR) can be expressed by \cite{idr1,idr2}
 
\begin{eqnarray} 
\frac{\textrm{Re}\,F_{+}(s)}{s}  \equiv
\frac{K}{s} + \frac{2s}{\pi}\,P\int_{s_o}^{\infty} 
ds' \left[\frac{1}{s'^2-s^2}\right]  
\frac{\textrm{Im}\,F_{+}(s')}{s'} , 
\label{ea4}
\end{eqnarray}

\begin{eqnarray} 
\frac{\textrm{Re}\,F_{-}(s)}{s} \equiv
  \frac{2}{\pi}\,P\int_{s_o}^{\infty} 
ds' \left[\frac{s'}{s'^2-s^2}\right]  
\frac{\textrm{Im}\,F_{-}(s')}{s'},
\label{ea5}
\end{eqnarray}
where $K$ is the \textit{subtraction constant} and $P$ denotes principal value.

On the other hand, for classes of functions of interest, IDR
can be replaced by derivative forms \cite{kn,ddr1,ddr2,ddr3,ddr4,ddr5,bks}, known as \textit{derivative dispersion relations}
(DDR). These may be more useful in some practical calculations,
as is the case here with the $\ln^{\gamma}(s/s_h)$ term for $\gamma$ real.
In these formulas, differentiation with
respect to the logarithm of the energy occurs in the argument of a
trigonometric operator, as in the
standard form deduced by Bronzan, Kane and Sukhatme
\textit{in the high-energy approximation} ($s_0 \rightarrow 0$ in equations (\ref{ea4}) and (\ref{ea5})) \cite{bks}:

\begin{eqnarray} 
\frac{\textrm{Re}\,F_{+}(s)}{s} =
\frac{K}{s} + 
\tan\left[\frac{\pi}{2}\frac{d}{d\ln s} \right] 
\frac{\textrm{Im}\,F_{+}(s)}{s}, 
\label{ea6}
\end{eqnarray}

\begin{eqnarray} 
\frac{\textrm{Re}\,F_{-}(s)}{s} =
\tan\left[\frac{\pi}{2}\left( 1 + \frac{d}{d\ln s}\right) \right]
\frac{\textrm{Im}\,F_{-}(s)}{s},
\label{ea7}
\end{eqnarray} 
corresponding, as before, to singly-subtracted DDR, under the high-energy approximation
(we shall return to this approximation in section A.2).

In order to implement the calculation, including the logarithm parametrizations,
we have used the operator expansion form introduced by
Kang and Nicolescu \cite{kn} (also discussed in \cite{am04}):

\begin{eqnarray} 
\fl \frac{\textrm{Re}\ F_{+}(s)}{s} = \frac{K}{s} +
\left[ \frac{\pi}{2} \frac{d}{d\ln s} + 
\frac{1}{3} \left(\frac{\pi}{2}\frac{d}{d \ln s}\right)^3 +
\frac{2}{15} \left(\frac{\pi}{2}\frac{d}{d \ln s}\right)^5 + \
.\ .\ .\ \right] \frac{\textrm{Im}\ F_{+}(s)}{s},
\label{ea8}
\end{eqnarray}

\begin{eqnarray} 
\frac{\textrm{Re}\ F_{-}(s)}{s}  &=& 
- \int \left\{ \frac{d}{d\ln s} \left[\cot \left( \frac{\pi}{2} 
\frac{d}{d\ln s} \right)\right]\frac{\textrm{Im}\ F_{-}(s)}{s} \right\} d\ln s \nonumber \\
&=&
- \frac{2}{\pi}\int \left\{ \left[ 1 - \frac{1}{3} \left(\frac{\pi}{2}\frac{d}{d \ln
s}\right)^2   \right. \right. \nonumber \\
&-& \left. \left. \frac{1}{45} \left(\frac{\pi}{2}\frac{d}{d \ln s}\right)^4
 - .\ .\ .\ \right] \frac{\textrm{Im}\ F_{-}(s)}{s} \right\} \, d \ln s.
\label{ea9}
\end{eqnarray}

Through these two expansions, with the parametrizations defined for
$\sigma_{\mathrm{tot}}(s)$, equations (\ref{e4}) - (\ref{e8}), the optical theorem (\ref{e1}), the $\rho$
definition (\ref{e3}) and the above equations (\ref{ea1}) - (\ref{ea3}), we obtain the
analytic results for $\rho(s)$ given by equations (\ref{e9}) - (\ref{e14}) in section \ref{s22}.

\subsection{Comments on the practical use of derivative dispersion relations}
\label{saa2}

For our purposes, it is important to note that equations (\ref{ea4}) and (\ref{ea5}) are not 
analytically equivalent to equations 
(\ref{ea6}) and (\ref{ea7}) due to the \textit{high-energy approximation} involved. For the functions 
of interest here, it is possible
to obtain an exact DDR result by taking into account the extremes at $s = s_0$  after
integration by parts. The result, however, introduces an infinite or double infinite
series, depending on the assumptions involved. The main point concerns the term associated 
with the lower limit of the primitive: assuming it to be zero (if the imaginary part of the amplitude
vanishes at the threshold $s_0 = 4m_p^2$), Cudell, Martynov and Selyugin have obtained a single series
\cite{cudell1,cudell2}
and without that assumption \'Avila and Menon have obtained a double infinite series \cite{am1,am2},
corresponding therefore to a general expression.
This last result can also be put in the form of a single series using sum rules of the incomplete
Gamma function, as demonstrated by Ferreira and Sesma \cite{fs} (see also \cite{fs2} for a recent
discussion on these representations and further results).

On the other hand, it is also possible to avoid the use of infinite series. The point is to take into account
the practical equivalence between the IDR (exact results) and the above DDR 
(with the high-energy approximation), once the subtraction constant is used as a free fit
parameter. This equivalence has been demonstrated by  \'Avila and Menon \cite{am04}
and also verified by other authors \cite{cudell1,cudell3}: the high-energy approximation is absorbed
by an effective subtraction constant.
Here, we have assumed this strategy, namely we treat $K$ as a free fit parameter. It is important to stress the main point involved:
for the functions of interest in amplitude analyses,
\textit{in order to obtain practical equivalence between IDR and DDR results, the subtraction constant
must be employed as a free fit parameter}. To take $K=0$ or to neglect it
does not guarantee the correct use of derivative relations.
In this respect, it is important to note that in the COMPETE analysis, the authors
refer to the use of DDR, but there is no information on the subtraction constant \cite{compete1,compete2}.

A second \textit{fundamental} aspect connected with the subtraction constant
demands also some comments. In both IDR and DDR, the subtraction
constant appears in the form $K/s$, suggesting that its influence (effect) is limited
to the low-energy region. That, however, is not the case in global fits to
$\sigma_{\mathrm{tot}}$ and $\rho$ data due to the nonlinear character of the data reduction:
$K$ as a free parameter is strongly correlated with all the other free parameters involved,
including those in the $\sigma_{HE}(s)$ contribution. This effect has been demonstrated
and discussed in \cite{alm01,alm03} in the cases of $s^{\epsilon}$ and $\ln^2s$ leading
contributions; it is also illustrated in our previous analyses with the
$\ln^{\gamma}s$ form. (See, for example,  table 6 in \cite{ms1}: the correlation
coefficient between $K$ and $\gamma$ is around 0.8.)

At last, it is well known that $\rho$ is in reality a free parameter in
fits to the differential cross section data in the region of the Coulomb-nuclear
interference. Therefore it does not have the same character of the total 
cross-section as an effective physical quantity. Moreover, the inclusion of the $\rho$
information in global fits to  $\sigma_{\mathrm{tot}}$ and $\rho$ data
constraints the rise of the total cross-section. This effect is also related
to the subtraction constant due to its correlation with all the fit parameters,
as discussed in \cite{fms2,ms1}.

Based on the above facts, we understand that individual fits to $ \sigma_{\mathrm{tot}}$ data,
together with checks on the corresponding predictions for $\rho(s)$ (using DDR), constitute
a more adequate procedure than to treat global fits. Despite our focus on the individual
fits, global fits are also treated as a complement in Appendix B and referred to in section
\ref{s4}.

\section{Global fits to total cross-section and $\rho$ data}
\label{sab}

\begin{figure}[pb]
\centering
\epsfig{file=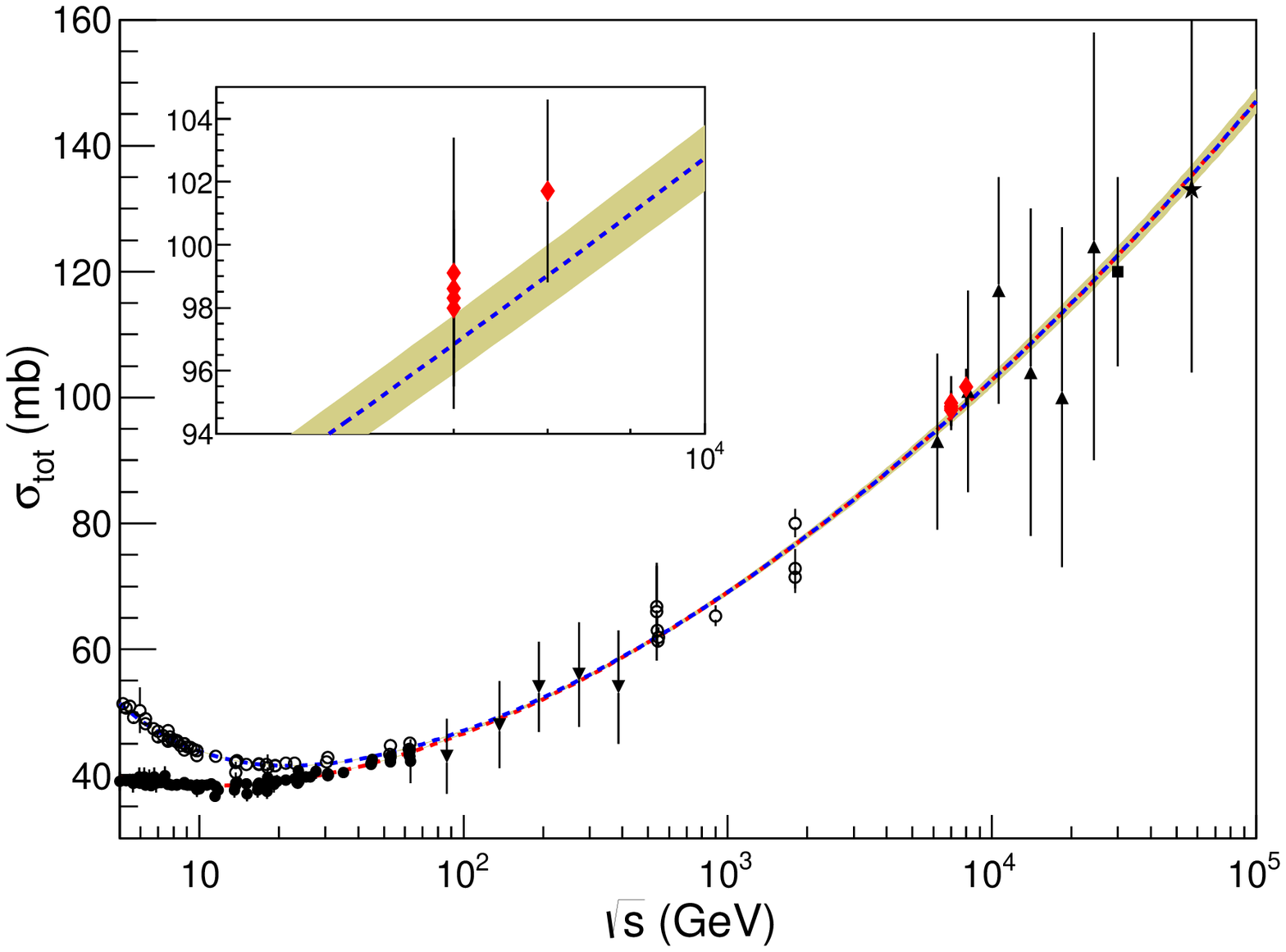,width=10cm,height=6cm}
\epsfig{file=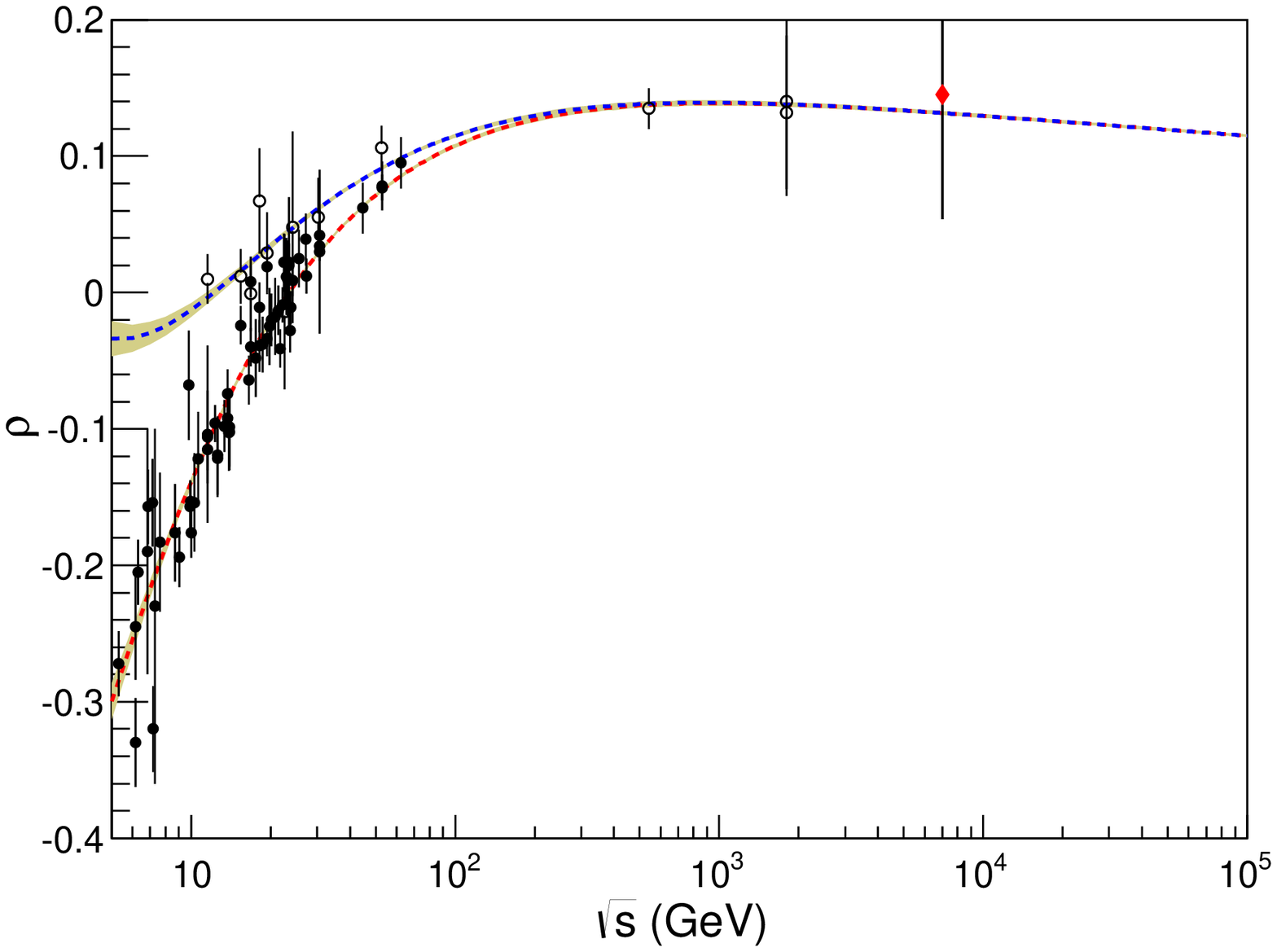,width=10cm,height=6cm}
\epsfig{file=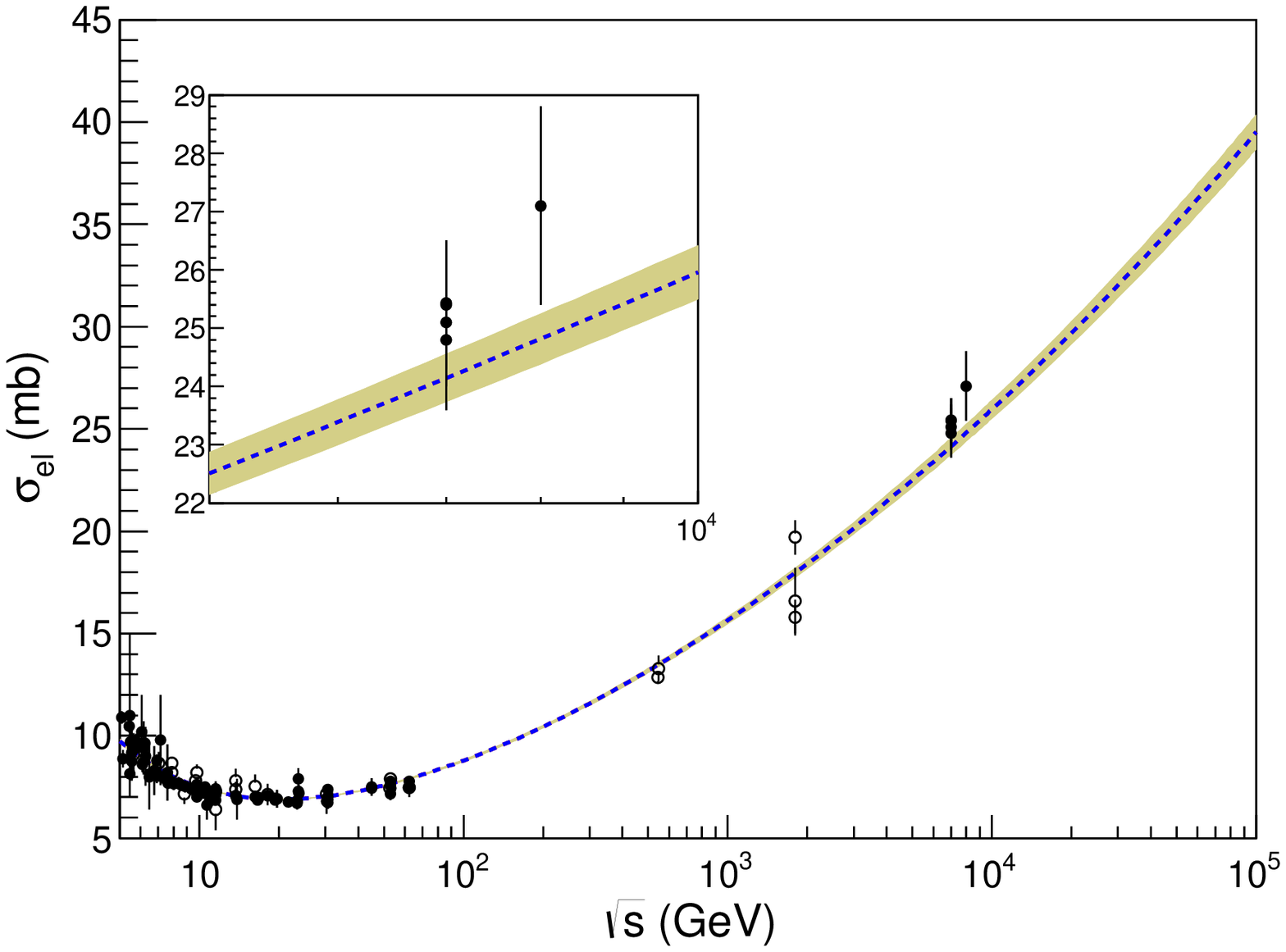,width=10cm,height=6cm}
\epsfig{file=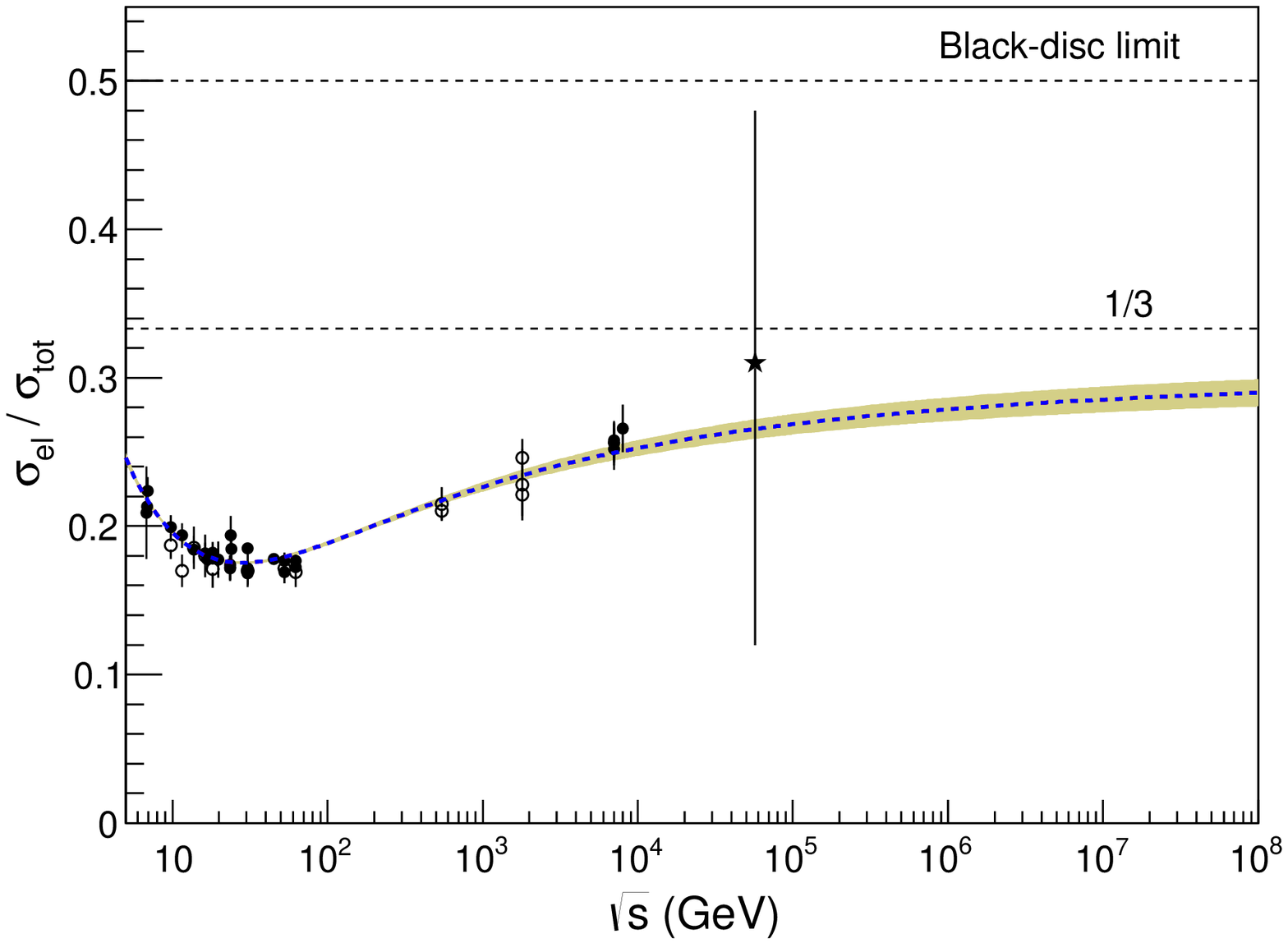,width=10cm,height=6cm}
\caption{Results of the global fits to $\sigma_{\mathrm{tot}}$ and $\rho$ data
with the $L2$ model, $s_h = 4m_p^2$ fixed and the extensions to the
elastic cross-section (table \ref{t8}, second and third columns).}
\label{f11}
\end{figure}

\begin{figure}[pb]
\centering
\epsfig{file=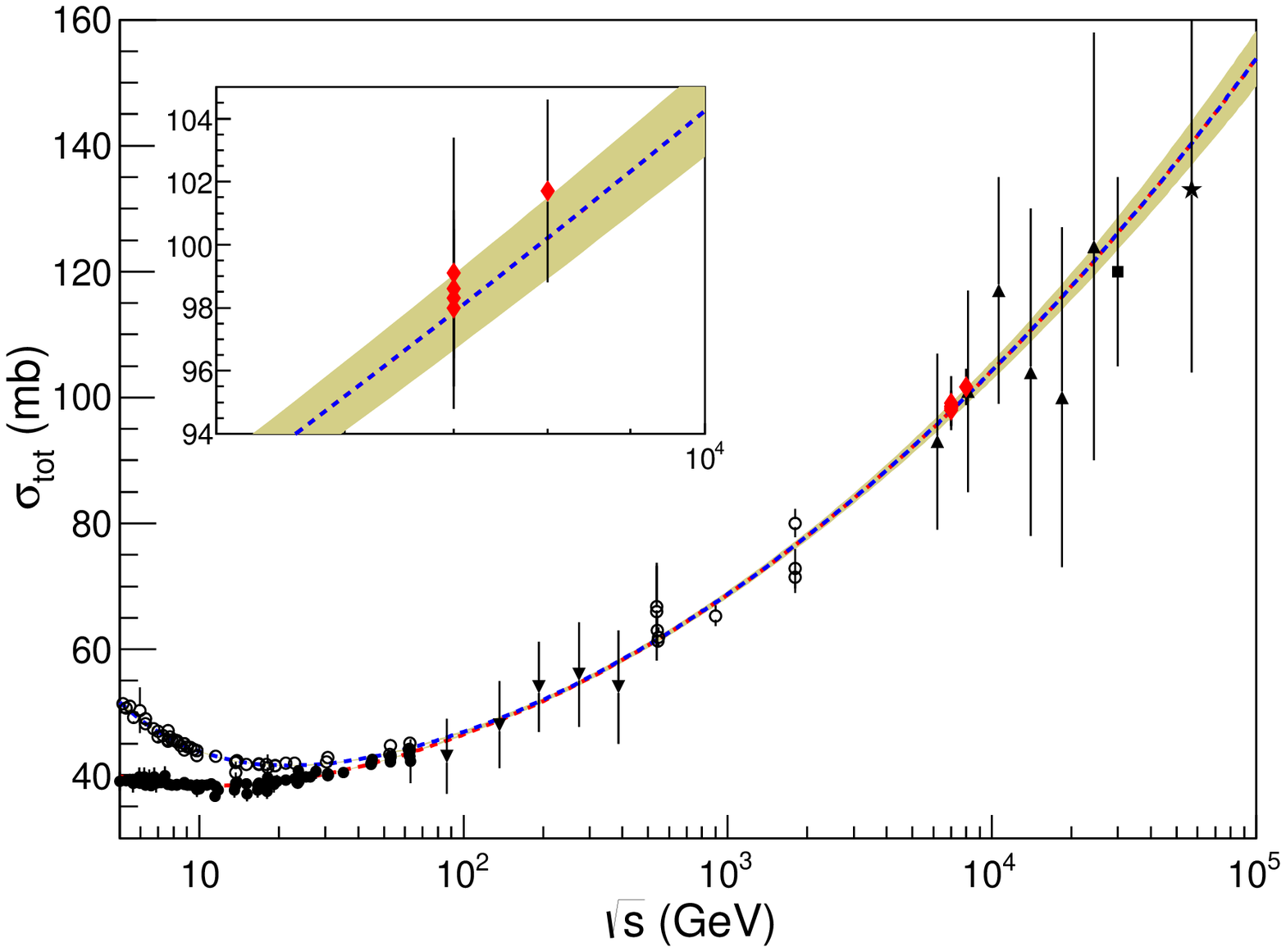,width=10cm,height=6cm}
\epsfig{file=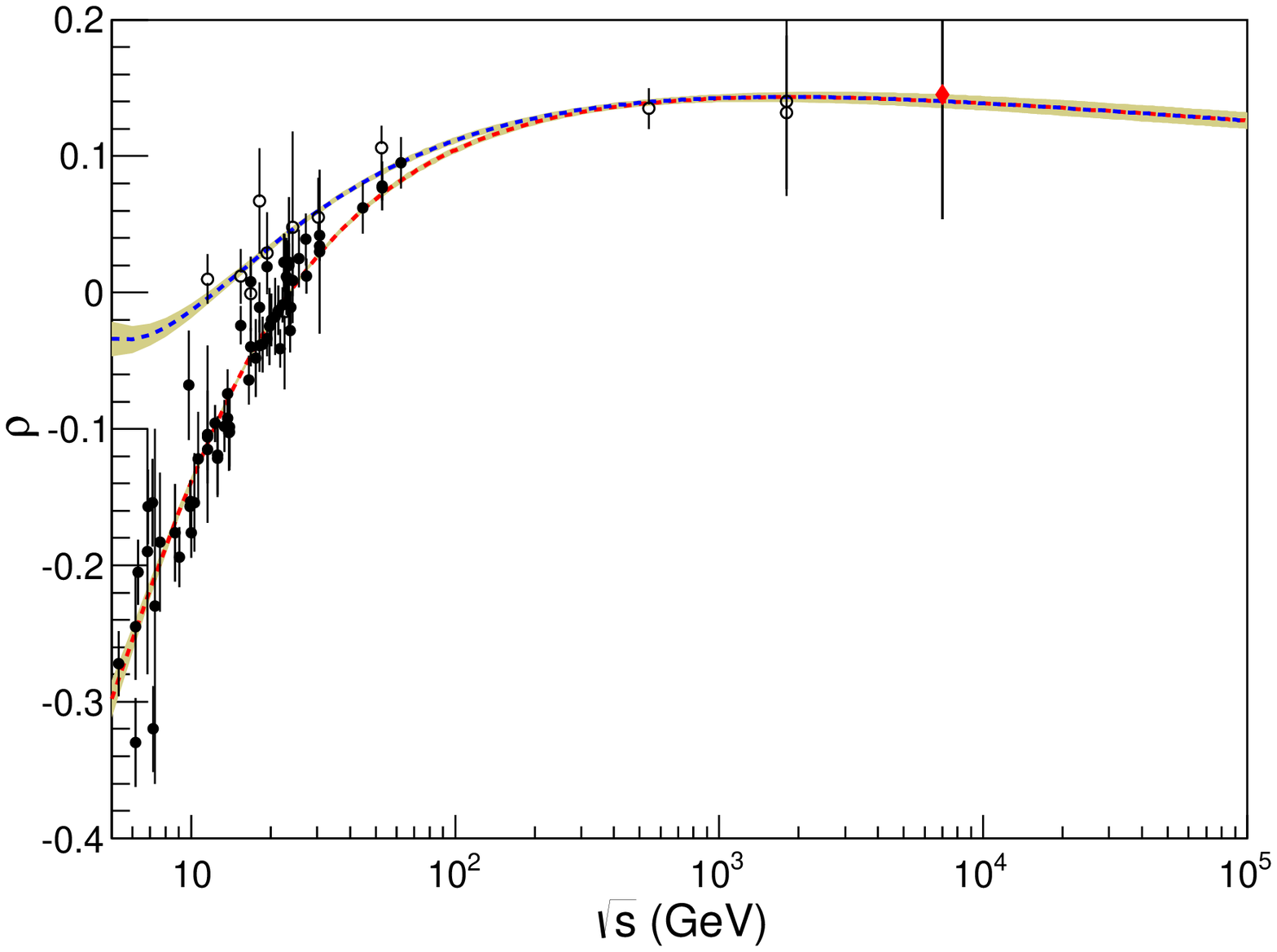,width=10cm,height=6cm}
\epsfig{file=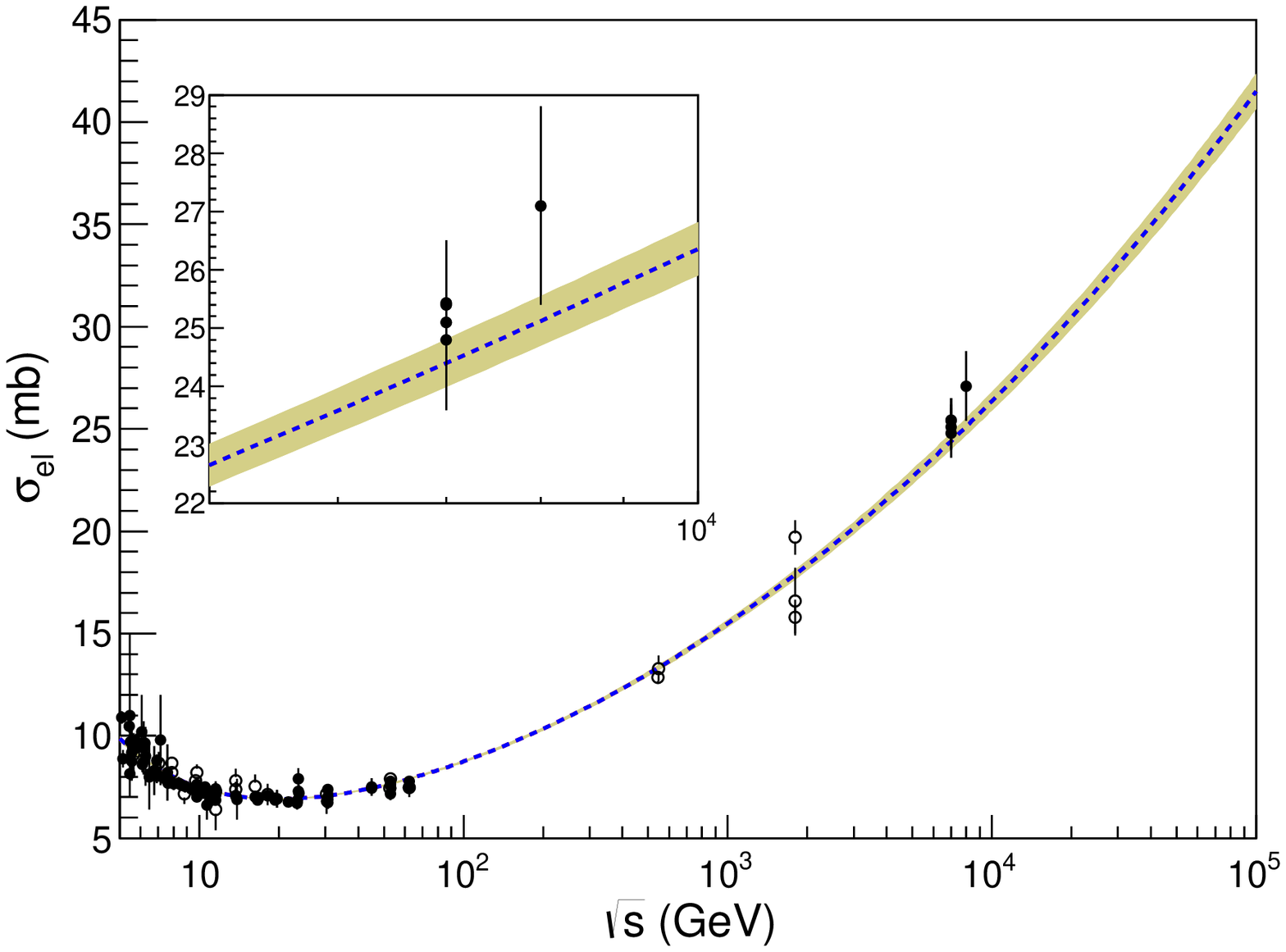,width=10cm,height=6cm}
\epsfig{file=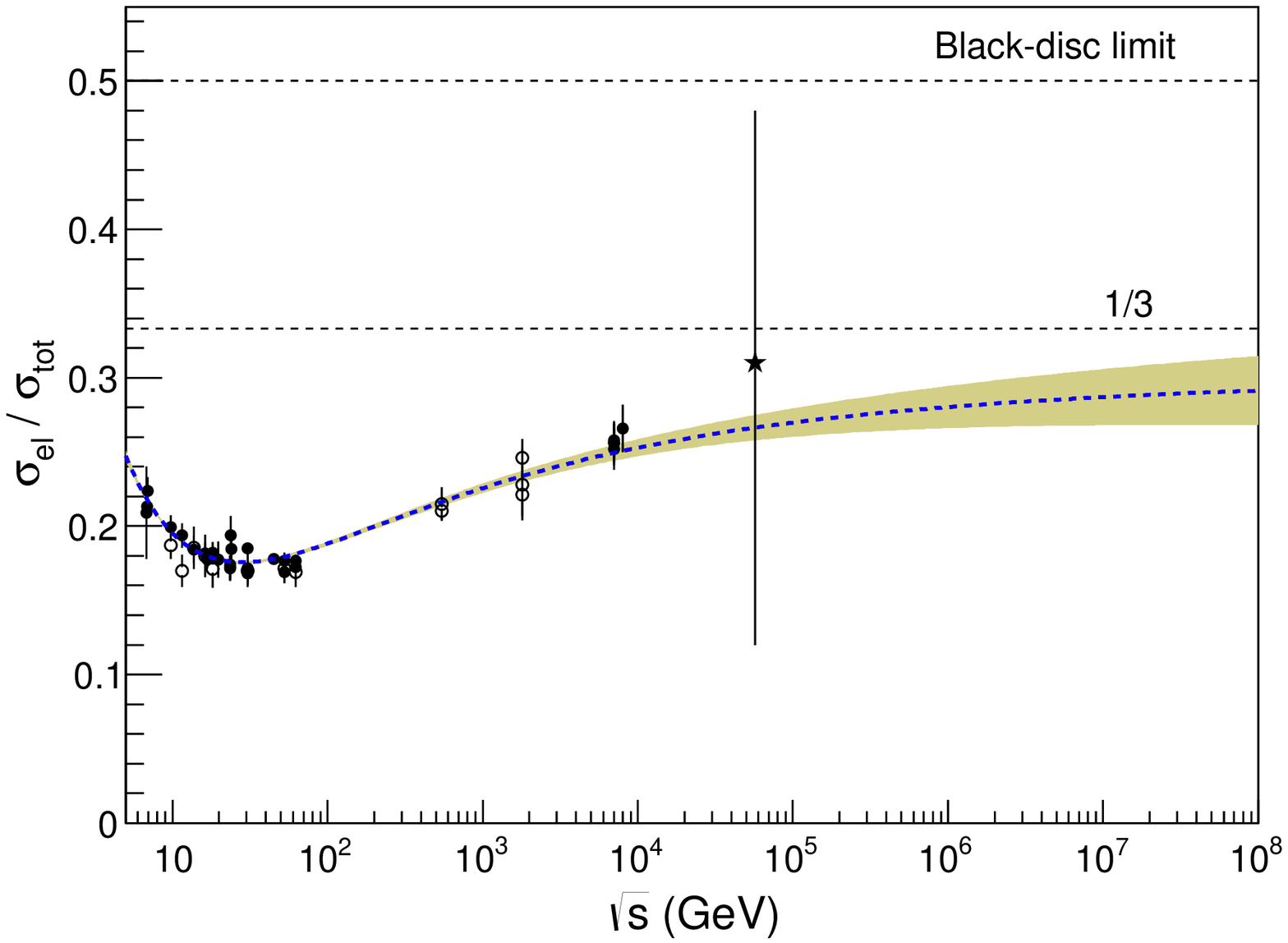,width=10cm,height=6cm}
\caption{Results of the global fits to $\sigma_{\mathrm{tot}}$ and $\rho$ data
with the $L\gamma$ model, $s_h = 4m_p^2$ fixed and the extensions to the
elastic cross-section (table \ref{t8}, fourth and fifth columns).}
\label{f12}
\end{figure}

In this appendix, we present the results of the global (simultaneous) fits
to $\sigma_{\mathrm{tot}}$ and $\rho$ data with the $L2$ and $L\gamma$ models 
in the case of $s_h = 4m_p^2$ fixed. As initial values, we have used once more the central values of
the COMPETE results (table \ref{t1}) and $K$ = 0.
The results, obtained through parametrizations (\ref{e4}), (\ref{e5}), (\ref{e7}) and (\ref{e8})
for $\sigma_{\mathrm{tot}}$ and 
(\ref{e9}), (\ref{e10}), (\ref{e12}), (\ref{e13}) and (\ref{e14}) for $\rho(s)$, 
are displayed in the second and fourth columns of table \ref{t8}
and figures \ref{f11} and \ref{f12}. The results of the extensions to $\sigma_{\mathrm{el}}$ data
are shown in the third and fifth columns of table \ref{t8} and the corresponding
asymptotic ratios $\sigma_{\mathrm{el}}$/$\sigma_{\mathrm{tot}}$ in table \ref{t9}.

\Table{\label{t8}Results from global fits to $\sigma_{\mathrm{tot}}$ and $\rho$ data and the extensions
to $\sigma_{\mathrm{el}}$ data with the $L2$ and $L\gamma$ models in the case of
$s_h = 4m_p^2$ fixed. Units as in table \ref{t1}.}
\begin{tabular}{c| c c| c  c}\hline
               & \multicolumn{2}{c|}{$L2$ model}        & \multicolumn{2}{c}{$L\gamma$ model} \\

               &     $\sigma_{\mathrm{tot}}$  &     $\sigma_{\mathrm{el}}$       &      $\sigma_{\mathrm{tot}}$  &  $\sigma_{\mathrm{el}}$\\\hline
  $a_1$        &   53.2 $\pm$ 2.1    &    27.0 $\pm$ 2.8       &    59.5 $\pm$ 5.8    &   32.4 $\pm$ 4.3\\
  $b_1$        &  0.400 $\pm$ 0.018  &   0.480 $\pm$ 0.038     &   0.505 $\pm$ 0.058  &  0.583 $\pm$ 0.042\\
  $a_2$        &   33.8 $\pm$ 2.0    &      0 (fixed)          &    34.1 $\pm$ 2.0    &      0 (fixed)\\
  $b_2$        &  0.545 $\pm$ 0.013  &      $-$                &   0.547 $\pm$ 0.013  &      $-$      \\
  $\alpha$     &  29.88 $\pm$ 0.46   &    3.68 $\pm$ 0.23      &    33.4 $\pm$ 1.4    &   4.70 $\pm$ 0.15\\
  $\beta$      & 0.2473 $\pm$ 0.0049 &  0.0756 $\pm$ 0.0022    &   0.124 $\pm$ 0.041  & 0.0382 $\pm$ 0.0010\\
  $\gamma$     &       2 (fixed)     &      2 (fixed)          &    2.23 $\pm$ 0.11   &    2.23 (fixed)\\
  $s_h$        &     3.521 (fixed)   &    3.521 (fixed)        &     3.521 (fixed)    &    3.521 (fixed)\\
  $K$          &   22.6 $\pm$ 6.9    &         $-$             &      40 $\pm$ 13     &         $-$\\\hline
  DOF          &         238         &         104             &          237         &         104\\
  $\chi^2$/DOF &         1.09        &         1.72            &          1.08        &         1.64\\
  $P(\chi^2)$  &        0.163        &  7.77$\times$10$^{-6}$  &         0.184        &  4.41$\times$10$^{-5}$\\
\hline
Figure         &    \ref{f11}        &       \ref{f11}         &    \ref{f12}         &       \ref{f12} \\
\hline
  \end{tabular}
\endTable

% \vspace{1.0cm}

\Table{\label{t9}Asymptotic results for the ratio $\sigma_{\mathrm{el}}/\sigma_{\mathrm{tot}}$, obtained from the global fits
to $\sigma_{\mathrm{tot}}$ and $\rho$ data, with the $L2$ and $L\gamma$ models for $s_h = 4m_p^2$ fixed and the extensions to $\sigma_{\mathrm{el}}$, equation (\ref{e15}).}
\begin{tabular}{c|c}\hline
      Model   &     Asymptotic  Ratio $\sigma_{\mathrm{el}}/\sigma_{\mathrm{tot}}$ \\
\hline
$L2$          &   0.306 $\pm$ 0.011 \\
$L\gamma$     &    0.31 $\pm$ 0.10 \\
\hline
\end{tabular}
\endTable

% \vspace{1.0cm}


\end{document}